\documentclass[conference]{IEEEtran}
\ifCLASSINFOpdf
\else
\fi

\usepackage{algorithmic,times}
\usepackage{graphicx}
\usepackage{booktabs}
\usepackage{blindtext, graphicx}
\usepackage{amsmath, amsthm}
\usepackage{amsfonts}
\usepackage{stfloats}
\usepackage{subfigure}
\usepackage{slashbox}
\usepackage{algorithm}
\usepackage{mathrsfs}
\usepackage{amssymb}

\usepackage{array}
\usepackage{multirow}
\usepackage{threeparttable}
\usepackage{color}

\begin{document}
\title{Interference Steering to Manage Interference}

%% author names and affiliations
%% use a multiple column layout for up to three different
%% affiliations
%\author{\IEEEauthorblockN{Michael Shell}
%\IEEEauthorblockA{School of Electrical and\\Computer Engineering\\
%Georgia Institute of Technology\\
%Atlanta, Georgia 30332--0250\\
%Email: http://www.michaelshell.org/contact.html}
%\and
%\IEEEauthorblockN{Homer Simpson}
%\IEEEauthorblockA{Twentieth Century Fox\\
%\and
%\IEEEauthorblockN{James Kirk\\ and Montgomery Scott}
%\IEEEauthorblockA{Starfleet Academy\\
%San Francisco, California 96678--2391\\
\author{\IEEEauthorblockN{Zhao Li$^{*}$, Fengjuan Guo$^{*}$, Kang G. Shin$^{\dagger}$, Yinghou Liu$^{*}$, Jia Liu$^{\ddagger}$}
\IEEEauthorblockA{$^{*}$State Key Laboratory of Integrated Service Networks, Xidian University\\
$^{\dagger}$The University of Michigan, USA\\
$^{\ddagger}$National Institute of Informatics, Japan\\
zli@xidian.edu.cn, fjguo@stu.xidian.edu.cn, kgshin@umich.edu, yhliu$\_$2017@stu.xidian.edu.cn, jliu@nii.ac.jp}}


%Telephone: (800) 555--1212, Fax: (888) 555--1212}


\maketitle

\begin{abstract}
To enable densely deployed base stations (BSs) or access points (APs)
to serve an increasing number of users and provide diverse mobile services,
we need to improve spectrum utilization in wireless communication networks.
Although spectral efficiency (SE) can be enhanced via smart and dynamic resource allocation,
interference has become a major impediment in improving SE.
There have been numerous interference management (IM) proposals at the
interfering transmitter or the victim transmitter/receiver separately or cooperatively.
Moreover, the existing IM schemes rely mainly on the use of channel state information (CSI).
However,  in some communication scenarios, the option to adjust the interferer is not available,
and, in the case of downlink transmission, it is always difficult or even impossible
for the victim receiver to acquire necessary information for IM.
%One can also exploit both CSI and interference information in IM design.

Based on the above observations, we first propose a novel IM technique,
called {\em interference steering} (IS).
%By making use of both CSI and data information w.r.t.~interference,
By making use of both CSI w.r.t.~and data carried in the interfering signal,
IS generates a signal to modify the spatial feature of the original interference,
so that the steered interference at the victim receiver is orthogonal to its intended signal.
We then apply IS to an infrastructure-based enterprise wireless local area network (WLAN) in which the same frequency
band is reused by adjacent basic service sets (BSSs) with overlapping areas.
With IS, multiple nearby APs could simultaneously transmit data on the same channel
to their mobile stations (STAs), thus enhancing spectrum reuse.
Our in-depth simulation results show that IS significantly improves
network SE over existing IM schemes.\end{abstract}


% \fi
\IEEEpeerreviewmaketitle

\section{Introduction}
Next-generation (a.k.a.~IMT-2020 or 5G) mobile wireless networks are characterized
by high data rate, high system capacity, and massive device connectivity [1].
Wireless networks circa 2020 are expected to be 1000x larger in capacity and
capable of connecting 100 billion devices [2]. In order to meet these requirements,
more than 1000MHz of new frequency bands should be identified to fill the spectrum
resource gap by 2020. Moreover, efficient spectrum utilization policies and techniques
should be developed to enable densely-populated BSSs or APs to serve an increasing
number of users and provide diverse mobile services.
However, interference will rise with the increase of spectrum utilization,
and must be well addressed so as to achieve high spectrum efficiency.
%======================================================
%EE We delete this part on May 1st, 2016 %%START HERE%%
%======================================================
%For example, cognitive radio (CR) [3] can be employed
%to allow unlicensed users to dynamically access the licensed band.
%In such a system, unlicensed users' timely detection of a licensed signal
%and immediate vacation of their spectrum are critical to avoid or minimize the interference to the licensed users' transmissions.
%Moreover, device-to-device (D2D) [4] and small cells [5] can be deployed on top of
%existing cellular systems to offload the macrocell and boost spectral efficiency (SE)
%via spatial reuse [6], but co-channel interference (CCI) between multiple concurrent
%transmissions occurs, degrading SE.
%Thus, effective IM mechanism is essential to improve SE.
%======================================================
%EE We delete this part on May 1st, 2016   %%END HERE%%
%======================================================

There have been numerous promising proposals to manage interference, which
can be classified into two types. The first is isolating mutually interfering transmissions
via resource partition, such as fractional frequency reuse (FFR), soft frequency
reuse (SFR) [3], enhanced inter-cell interference coordination (eICIC) [4], etc.
However, these mechanisms may cause degradation of spectrum efficiency.
The second type is employing various signal processing techniques, such as zero-forcing
beamforming (ZFBF) [5], zero-forcing (ZF) reception [6], coordinate multi-point (CoMP) [7],
interference cancellation (IC) [8], interference alignment (IA) [9],
interference neutralization (IN) [10-14], etc., to support concurrent transmission of
multiple interfering signals.

%Of these signal processing methods,
%IA has been under development in recent years [13-16].
%Its principle is to confine all interferences to a subspace of minimal dimensions at the
%receiver so as to maximize the available dimensions for the intended signals.
%Via preprocessing at the transmitter side, multiple interfering signals are mapped into
%a finite subspace, so that desired signal(s) may be sent through a subspace without attenuation.
%The authors of [13]
%showed that the feasibility of IA is highly dependent on system parameters,
%such as the numbers of transmitters and receivers,
%configuration of transmitting and receiving antennas, etc.
%Opportunistic IA was proposed in [14] for a large number of users
%to harvest the multiuser diversity so as to facilitate the implementation of IA.
%IA-based coordinated beamforming was proposed in [15] to improve the downlink performance of
%multiple cell-edge users in multi-user multiple-input multiple-output (MU-MIMO) systems.
%IA-based uplink IM for two-tier cellular systems was devised in [16].

%Interference can be not only aligned but also canceled or partially canceled through multiple
%paths, which are referred to as {\em interference neutralization} (IN) [14-18].
Of these signal processing methods, IN has been under development in recent years [10-14].
It is a new IM mechanism found from, and inherent in interference networks with relays [11-12].
IN strives to combine signals arriving via various paths in such a way
that the interfering signals are canceled while preserving the desired signals [13].
It can be regarded as a distributed zero-forcing of interference before
the interfering signal reaches the undesired destination [14].
The authors of [13] constructed a linear distributed IN that encodes signals
in both space and time for separate multiuser uplink-downlink two-way communications.
In [14], an aligned IN was proposed in a multi-hop interference network formed
by concatenation of two two-user interference channels.
It provides a way to align interference terms over each hop in a manner that
allows them to be canceled over the air at the last hop.

Note, however, that none of these existing IM methods are free of cost. For example,
the CSI is required for implementing IA, ZF and ZFBF,
whereas both CSI w.r.t.~and data carried in the interfering signal are exploited
for IC, CoMP and IN.
By shaping a transmit beam using ZFBF, IA or CoMP, the adjusted signal will be attenuated;
a ZF-based filter can be adopted to nullify interference at the expense of degrading the desired
signal power somewhat; and for IN, an interfering signal is duplicated to neutralize
the interference at the cost of additional transmit power consumption.
Moreover, ZFBF and ZF reception require multiple antennas or degrees of freedom (DoFs)
at the transmitter (Tx) and the receiver (Rx), respectively,
and for IA, both ends of the communication link should be equipped with multiple antennas.
The DoF requirements of these methods are determined by the signal dimensions,
i.e., for ZFBF and ZF reception, each interfering signal component consumes one DoF,
whereas for IA, at least one additional DoF should be provided to place the aligned interference.
Moreover, IA is not applicable if multiple interferences are from an identical transmitter.
This is because if the interfering signals originated from the same source are aligned in one direction,
they will also overlap with each other at their intended receivers, thus becoming indistinguishable.
With IN, since interference(s) can be neutralized over the air,
no additional receiver-side DoF is required to cancel interference,
thus becoming free from the aforementioned limitations of ZF reception and IA.
\begin{table*}[htb!]
%\vspace*{-0.1in}
%\centering
\begin{center}
\caption{Comparison of IM methods.}
%\footnotesize
%\small
%\scriptsize
\begin{tabular}{|l|c|c|c|c|c|c|c|}\hline
\backslashbox[28mm]{Feature}{Method} & ZFBF & ZF & CoMP & IC & IA & IN & IS \\\hline
Tx beam adjustment &$\circ$ &$\times$ &$\circ$ &$\times$ &$\circ$ &$\times$ &$\times$ \\\hline
Rx beam adjustment &$\times$ &$\circ$ &$\times$ &$\times$ &$\times$ &$\times$ &$\times$ \\\hline
Signal attenuation &$\circ$ &$\circ$ &$\circ$ &$\times$ &$\circ$ &$\times$ &$\times$ \\\hline
Tx power cost &$\times$ &$\times$ &$\times$ &$\times$ &$\times$ &$\circ$ &$\circ$ \\\hline
Tx side CSI exchange &$\circ$ &$\times$ &$\circ$ &$\times$ &$\circ$ &$\circ$ &$\circ$ \\\hline
Tx side data exchange &$\times$ &$\times$ &$\circ$ &$\times$ &$\times$ &$\circ$ &$\circ$ \\\hline
Rx side CSI exchange &$\times$ &$\circ$ &$\times$ &$\circ$ &$\times$ &$\times$ &$\times$ \\\hline
Rx side data exchange &$\times$ &$\times$ &$\times$ &$\circ$ &$\times$ &$\times$ &$\times$ \\\hline
Tx side multi-antenna &$\circ$ &$\times$ &$\circ$ &$\times$ &$\circ$ &$\circ$ &$\circ$ \\\hline
Rx side multi-antenna &$\times$ &$\circ$ &$\times$ &$\times$ &$\circ$ &$\times$ &$\circ$ \\\hline
Symbol-level synchrony &$\times$ &$\times$ &$\circ$ &$\circ$ &$\times$ &$\circ$ &$\circ$ \\\hline
\end{tabular}
\end{center}
\vspace*{-0.1in}
\end{table*}

Table~I compares some signal-processing-based IM,
where the symbols $\circ$ and $\times$ indicate
having and not having the corresponding feature, respectively.
Tx beam indicates data transmission from various sources to their corresponding
receivers. Rx beam means the direction of the receive filter's main lobe.
Either Tx or Rx beam adjustment will cause effective signal power loss.
Tx/Rx-side CSI exchange indicates that the transmitter/receiver needs to acquire
CSI from all Rxs/Txs, including intended and unintended Rxs/Txs, to itself.
In practice, ZFBF and IA are implemented at the interfering transmitter,
ZF is at the victim receiver, CoMP and IC are realized cooperatively
at the Tx and Rx, respectively. When the interfering Tx is unwilling to perform IM,
or collaboration between the interferer and the victim cannot be established,
all methods based on such collaboration won't be applicable.

The above discussion implies the need for a new solution to the interference problem
despite the numerous IM schemes proposed thus far.
Unlike those methods requiring implementation at either the interfering
Tx (for IA, ZFBF) or the victim Rx (for ZF), we need the victim Tx-side interference
management for the following reasons.
First, due to the equal or higher priority of interfering transmission over the victim,
the interferer may not be amenable to implementing IM, especially when it incurs
some performance loss. For example, in a heterogeneous cellular network (HCN)
where small cells are deployed on top of a macrocell to improve the capacity and
coverage of existing cellular systems, the macro base station may interfere with
small cell users. Due to the small cells' subordinate nature, it is not practical to
modify macrocell's transmissions for the small victim cell users.
Second, in some communication scenarios such as a downlink transmission,
acquiring information for IM (e.g., interferer identification)
is always difficult or even impossible for the victim Rx,
especially when there are multiple interferers.
% CC We add the following sentence so as to emphasize the unavailability of existing IM methods.
So,  in the above situations,
the approaches listed in Table~I (except for IN) are not applicable, and hence a new IM method is called for.
We therefore focus on two aspects --- the available information we can utilize
and the cost of IM --- in the development of novel IM.
As shown in Table~I, most existing methods exploit CSI while
ignoring the data information that interference signal carries.
Although IN is a victim Tx-based IM implementation which makes use of the interferer's data
information, to the best of our knowledge, no existing IN schemes account for the power cost.
That is, under a transmit power constraint, the more power consumed for IN,
the less available for the intended signal's transmission.
Especially when interference at the victim Rx is relatively strong,
there might not be sufficient power to generate a required neutralization signal,
thus making IN infeasible.
The features of the proposed IS is similar to that of IN except for the multi-antenna requirement at the Rx side.
The details can be seen in the following sections.

By recognizing that interference can be not only neutralized
but also steered in a particular direction,
we first propose a novel IM method, called {\em interference steering} (IS).
With IS, a duplicated interfering signal is generated to modify the spatial feature
of the original interference observed by the victim Rx,
thus enabling interference-free data transmission.
Then, we discuss the application of IS to an enterprise WLAN,
which has become a pervasive and essential part of our professional life,
and will be an increasingly important in future.
\begin{figure*}[!htb]
\vspace*{-0.1in}
\centering
\begin{minipage}{1\columnwidth}
\centering
\includegraphics[width=0.80\columnwidth, height = 0.50\columnwidth]{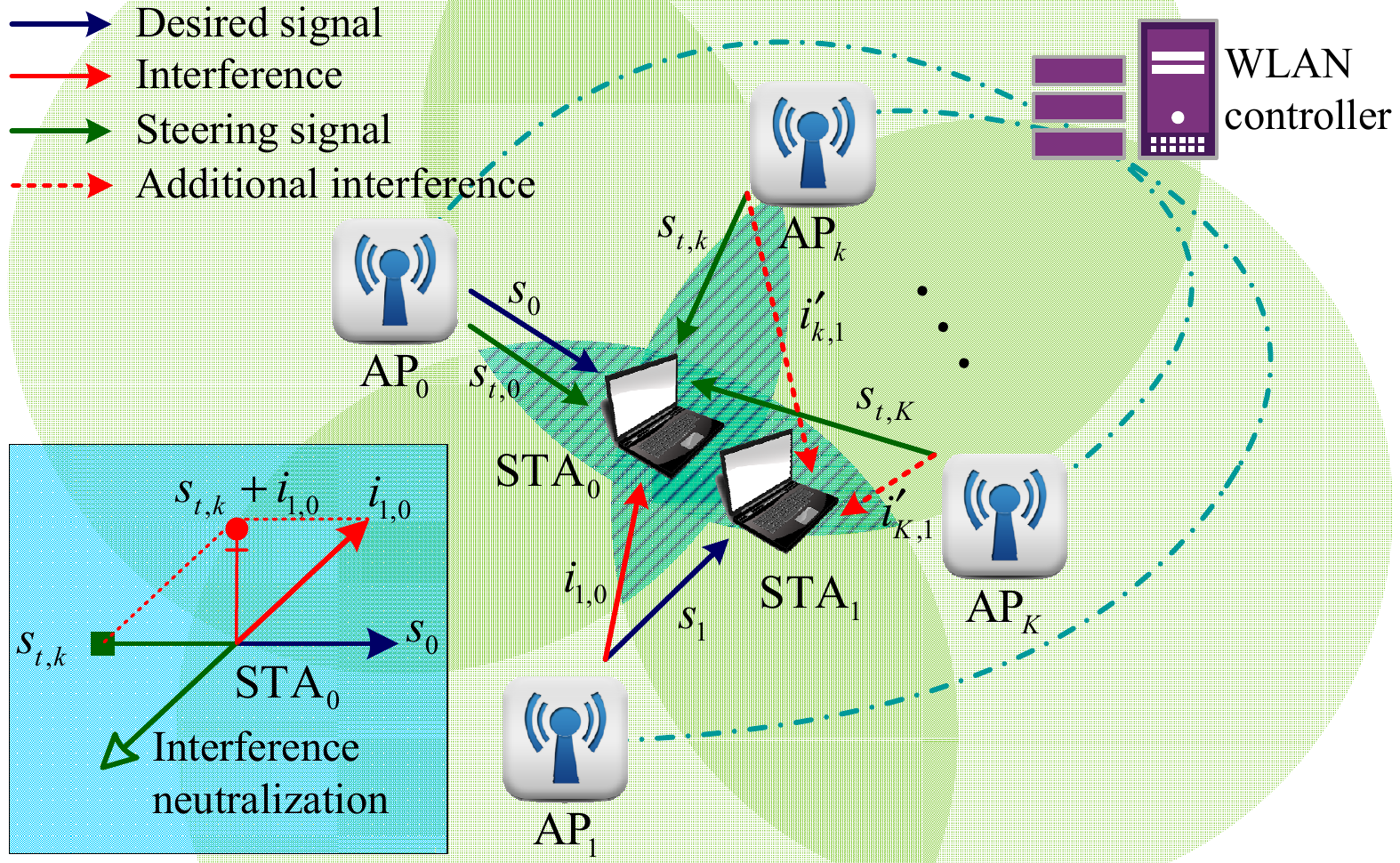}
\vspace{-5pt}
\caption{System model.}
\label{fig: figure one}
\end{minipage}
\hfill
\begin{minipage}{1\columnwidth}
\centering
\includegraphics[width=0.75\columnwidth, height = 0.50\columnwidth]{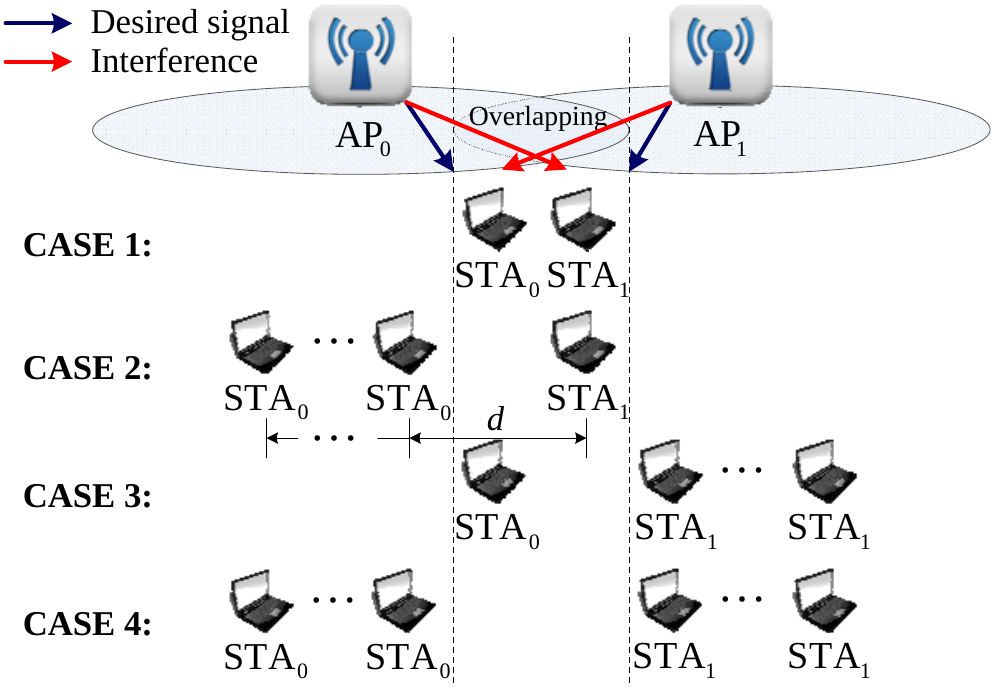}
\vspace{-5pt}
\caption{Interference scenarios.}
\label{fig: figure two}
\end{minipage}
\vspace*{-0.15in}
\end{figure*}

The main contributions of this paper are two-fold:
\begin{itemize}
%\vspace*{-0.05in}
\item Proposal of a novel IM scheme called {\em interference steering} (IS).
By generating a steering signal, the original interference imposed on the victim Rx
is steered to the orthogonal direction w.r.t.~the desired signal,
hence achieving interference-free transmission.
IS consumes less power than IN but requires an additional DoF at the victim Rx.
Moreover, IS can also subsume IN as a special case, thus becoming more general.
%\vspace*{-0.05in}
\item Discussion of the application of IS in an enterprise WLAN
where the same frequency band is reused by adjacent BSSs with overlapping areas.
A random network topology and arbitrary number of interferences are considered.
With the proposed mechanism, interference to cell-edge users can be mitigated,
allowing nearby APs to transmit to their associated STAs on the same channel
simultaneously and hence improving the system's SE.
%\vspace*{-0.05in}
\end{itemize}

%CCC Introduce organization of paper here.
%======================================================
%EE We delete this part on May 1st, 2016 %%START HERE%%
%======================================================
The rest of this paper is organized as follows. Section
II describes the system model, while Section III presents
the design of interference steering.
In Section IV, the application of IS
in enterprise WLANs is detailed. Section V evaluates the
proposed mechanism. Finally, Section VI concludes the paper.
%======================================================
%EE We delete this part on May 1st, 2016 %%END HERE%%
%======================================================

Throughout this paper, we use the following notations. The set of complex numbers
is denoted as $\mathbb{C}$, while vectors and matrices are represented by bold
lower-case and upper-case letters, respectively. Let $\mathbf{X}^{H}$, $\mathbf{X}^{T}$
and $\mathbf{X}^{-1}$ denote the Hermitian, transpose
%conjugate
and inverse of matrix $\mathbf{X}$.
$\|\cdot\|$ indicates the Euclidean norm. $E(\cdot)$ denotes statistical expectation.
$\langle\mathbf{a}, \mathbf{b}\rangle$ represents the inner product of two vectors.

\section{System model}

Consider downlink communication in an infrastructure-based enterprise WLAN
in which coverage areas of APs often overlap. For example, approximately 15\% more APs
(resulting in overlapping coverage) are required for wireless voice communications so as to
achieve an acceptable reception power level [15].
Although data-only communications may not require such a large amount of overlap,
in order to achieve seamless coverage, coverage overlap is inevitable.
For example, Fig.~1 shows $K$ adjacent BSSs with overlapping areas.
All APs are assumed to have the same transmit power, $P_{T}$, and connected to a central WLAN
controller so that downlink transmissions from APs to their clients/STAs are synchronized.
%% CC We add this description to explain how we can acquire the data information carried by the interfering signal
%% CC so that IS can be achieved.
%\textcolor{blue}{
%Since in future WLANs, more processing is expected to move from APs to the controller,
%i.e., an AP may only be responsible for generating radio frequency (RF) signals
%by following the instructions sent from the controller.
%The controller maintains the association status between the corresponding
%APs and STAs, the transmitting status of each AP, and the data information to be transmitted.
%By utilizing such information, an IM solution can be obtained.}
Although multiple STAs may be in the coverage area of an AP,
only one STA is served at a time by its associated AP via one frequency channel and
each STA is associated with one AP at a time.
For clarity of presentation, we show only 2 edge STAs: STA$_0$ is associated with AP$_0$
and STA$_1$ with AP$_1$.
We assume APs and STAs each are equipped with $N_t$ and $N_r$ antennas, respectively.
Since mobile stations/devices are subject to severer restrictions such as cost and hardware
than an AP, we assume $N_{t}\geq N_{r}>1$.
% 2016.1.12 delete
%Without loss of generality, we can set $N_t=N_r=2$;
%CCC Say why?
%This can be easily extended to the case of $N_{t}\geq N_{r}$.
$\mathbf{H}_{mn}\in\mathbb{C}^{N_{r}\times N_{t}}$ denotes the spatial channel
from AP$_m$ to STA$_n$. We employ a spatially uncorrelated
Rayleigh flat fading channel model so that the elements of
$\mathbf{H}_{mn}$ are modeled as independent and identically distributed
zero-mean unit-variance complex Gaussian random variables.
All users experience block fading, i.e., channel parameters
in a block consisting of several successive transmission cycles remain
constant in the block and vary randomly between blocks.
Let $s_m$, $m\in\{1,\cdots,K\}$, be the desired signal
from AP$_m$ to its currently serving client STA$_m$, and
$i_{m,n}$ ($n\in\{1,\cdots,K\}$) be the interference from AP$_m$ to STA$_n$.
$s_{t,m}$ is the steering signal generated by AP$_m$, while
$i'_{m,n}$ is the additional interference caused by the steering signal $s_{t,m}$ to STA$_n$.
%\begin{figure}[ht]
%%\begin{figure}[!htb]
%\graphicspath{/fig}
%\centering
%\includegraphics[width=0.4\textwidth]{fig/figure01.pdf}
%\caption{System model.}
%\label{fig: figure one}
%%\end{figure}
%\vspace*{-0.1in}
%\end{figure}

According to the existing IEEE 802.11 protocols,
a STA scan RF channels to search for beacons
advertising the presence of nearby APs.
When a radio receives a beacon frame, it acquires information about the
capability and configuration of the corresponding network,
and then lists available, eligible APs.
Based on the number of available APs in the list, the STA determines
whether it is an edge-user \footnote{We use the term \textit{edge} to indicate the status of a STA that can hear from more than one AP,
rather than the geographical location of the STA.} or not.
The STA also estimates CSI based on the received null data packet (NDP)
announcements from nearby APs [16]. In this system model, APs with overlapping
areas should be scheduled by the central controller so that their NDP
transmissions are serialized (to avoid collision).
By analyzing the transmitter address field [17] in the NDP
announcement, the STA can differentiate CSI from different APs.
In general, a non-edge STA estimates and then reports CSI to its associated AP,
while an edge-STA form CSI from its associated AP as well as adjacent APs
as a vector and sends the vector to its associated AP.
The AP then delivers the CSI or CSI vector to the central controller.
%CCC You can write this as an algorithm in pseudo code

%======================================================
%EE We delete this part on May 1st, 2016 %%START HERE%%
%======================================================
%Via spectrum planning, adjacent BSSs are allocated mutually orthogonal frequency channels,
%but this will lead to low overall spectral utilization.
%To improve spectrum utilization, we allow adjacent BSSs to share the same frequency band,
%i.e., frequency reuse. Furthermore,
%======================================================
%EE We delete this part on May 1st, 2016 %%START HERE%%
%======================================================
Based on WLAN protocols, two adjacent BSSs contend
for the same channel and cannot operate on the same channel simultaneously. However,
for the purpose of SE enhancement, we may reuse spectrum more aggressively
with effective interference management.
In such a case, co-channel interference (CCI) may occur, and a STA should thus be able to detect collision caused by CCI,
i.e., distinguishing collision from fading individually [18] or cooperatively [19] with its associated AP.
Fig.~2 illustrates common interference scenarios in WLANs.
%\begin{figure}[ht]
%%\begin{figure}[!htb]
%\graphicspath{/fig}
%\centering
%\includegraphics[width=0.35\textwidth]{fig/figure02.pdf}
%\caption{Interference scenarios.}
%\label{fig: figure two}
%%\end{figure}
%\vspace*{-0.1in}
%\end{figure}
The figure shows downlink transmissions from AP$_{0}$ to STA$_{0}$ and from AP$_{1}$
to STA$_{1}$,  initiate at time $t_{0}$ and $t_{1}$, respectively.
This simplified discussion can be readily extended to the case of
more APs and STAs as in real WLANs (see Sections III and IV).
Without loss of generality, we let $t_{1}\geq t_{0}$.
Both AP and STA follow the listen-before-talk (LBT) rule,
and channel reciprocity holds, i.e., if an AP can detect signal from an unassociated STA,
its transmission signal will interfere with the STA's reception.
Let $d$ be the distance between STA$_{0}$ and STA$_{1}$.
For simplicity, we assume inter-AP distance is large enough for APs
unable to hear each other.

Let's consider the downlink transmissions from AP$_{0}$ to STA$_{0}$ and from AP$_{1}$
to STA$_{1}$ as an example that initiate transmission at time $t_{0}$ and $t_{1}$ respectively, and
without loss of generality, we investigate the channel occupancy status observed by BSS$_1$ and
use a general expression $[\text{STA}_{0}\rightarrow \text{AP}_{1},\text{AP}_{0}
\rightarrow \text{STA}_{1}]$ to represent the reachability between unassociated
AP -- STA pairs. For example, if AP$_1$ overhears STA$_0$, then the first bit
in the above expression is $1$ else $0$.
Since the reachability between STAs doesn't affect the downlink interference,
the link $\text{STA}_{0}\rightarrow \text{STA}_{1}$ is omitted.
This simplified discussion can be readily extended to the case of
more APs and STAs as in real WLANs (see Sections III and IV).
Without loss of generality, we let $t_{1}\geq t_{0}$.
Both AP and STA follow the listen-before-talk (LBT) rule,
and channel reciprocity holds, i.e., if an AP can detect signal from an unassociated STA,
its transmission signal will interfere with the STA's reception.
For simplicity, we assume inter-AP distance is large enough for APs can ignore signals from each other. %unable to hear each other.}
So, the reachability in case 1 of Fig.~2 is $[1,1]$,
i.e., mutual interference occurs between two adjacent BSSs.
We can simply block one of the transmissions or employ a coordinated multi-point
(CoMP) [20] to serve two STAs with both APs simultaneously,
but CoMP requires modification of the interferer's transmission.
As for the second situation, the reachability status is $[0,1]$, and hence
AP$_1$ is allowed to initiate its transmission to STA$_1$ by employing an effective
IM method to protect STA$_1$'s reception. In case 3, the reachability status is $[1,0]$.
Since STA$_0$ is not associated with AP$_1$ and all BSSs are equal,
AP$_1$ can transmit data to STA$_1$ regardless of the ongoing transmission
from AP$_0$ to STA$_0$.
Then, IM should be employed to guarantee STA$_0$'s reception.
CoMP is not applicable for cases 2 and 3, since not all STAs can hear from both APs.
If we don't allow AP$_1$ to transmit to STA$_1$, an activated edge-STA
will block transmissions on the same frequency channel in all its adjacent BSSs
whose AP/STA is exposed to the interferer's signal radiation.
This is undesirable for the network's requirement of high SE and system capacity.
In the last situation, the reachability status is $[0,0]$, so two interference-free
transmissions can be established concurrently.
Based on the above discussion, we focus on IM for cases 2 and 3,
i.e., edge-STA in case of asymmetric interference.
%In case 1, STA$_1$ can overhear AP$_0$ so that
%it will not send RTS/CTS to AP$_{1}$, hence no CCI occurs.
%In this situation, AP$_1$ and STA$_1$ may hear the signal sent from STA$_0$ too,
%however, since STA$_0$ is not associated with AP$_1$,
%this will not impede the establishment of transmission from AP$_1$ to STA$_1$.
%In case 2, various $d$ is considered.
%When $d$ is large enough, STA$_0$ will become hidden terminal w.r.t. STA$_1$.
%Then neither STA$_1$ nor AP$_1$ can hear from STA$_0$.
%However, STA$_1$ can detect signal from AP$_0$, hence no transmission from AP$_1$ to STA$_1$ is established
%and no CCI is imposed onto STA$_0$.
%As for the third case, STA$_{0}$, a hidden terminal w.r.t. STA$_1$, is in the overlapping area of both APs.
%Since STA$_{1}$ is not covered by AP$_0$,
%the transmission from AP$_1$ to STA$_1$ can be established.
%Then STA$_0$ will be disturbed by AP$_1$.

Since a STA is capable of detecting collision, upon sensing a conflict, the victim STA
reports to its associated AP, the latter (victim Tx) asks the WLAN controller for assistance.
The controller checks its database and finds a solution for the requestor.
The above procedure is in accordance with the centralized management framework
proposed in 5G [2], where a high-level node manages all the information.
% CC We add this description to explain how we can acquire the data information carried by the interfering signal
% CC so that IS can be achieved.
Specifically, the controller maintains the association status between the corresponding
APs and STAs, the transmitting status of each AP, and the data information to be transmitted.
By utilizing such information, an IS solution can be obtained (see Section IV for details).
To limit the system overhead, only edge-STAs can ask their associated APs for IM.

One should note that although we take WLAN as an example to design our mechanism,
other types of network as long as they are featured as
1) direct/indirect\footnote{\textit{Direct collaboration} refers to direct information exchange between Txs, %[refer to our DIN paper],
whereas \textit{indirect} indicates that the cooperation is achieved via a central control node, e.g., WLAN controller in Fig.~1.}
collaboration between the interfering Tx and victim Tx is available,
and 2) the interference topology is asymmetric, our scheme is still applicable.

\section{Design of Interference Steering}

We present the signal processing procedure of {\em interference steering} (IS)
by exploiting both CSI w.r.t.~and data carried in the interference(s).
IS generates a steering signal to modify the interference's spatial feature,
so that the original interference is steered to the orthogonal direction of
the desired signal observed by the victim Rx.
In what follows, we first describe the basic design in terms of a single
interference and then discuss multi-interferences scenario and cooperative IS.

\subsection{Signal Processing of IS}

We assume beamforming (BF) is employed for each downlink communication.
As shown in Fig.~1, consider the transmission from AP$_0$ to STA$_0$ as an example
which is corrupted by the signal sent from AP$_1$,
the received signal at STA$_0$ with IS can be expressed as
%\vspace*{-0.05in}
\begin{equation}
%\vspace*{-0.05in}
\begin{aligned}
\mathbf{y}_{0}&=\sqrt{P_{T}-P^{s_t}_{OH}}\mathbf{H}_{0}\mathbf{p}_{0}x_{0}+\sqrt{{P}_{T}}
\mathbf{H}_{10}\mathbf{p}_{1}x_{1}\\
&+\sqrt{P^{s_t}_{k}}\mathbf{H}_{k0}\mathbf{p}^{s_t}_{k}x_{k}^{s_t}+\mathbf{z}_{0}
\end{aligned}.
%\vspace*{-0.05in}
\end{equation}

The first three items on the right hand side (RHS) of Eq.~(1)
denote the desired signal from AP$_0$,
interference from AP$_1$, and steering signal from AP$_k$ where $k\in\{1,\cdots,K\}$, respectively.
$\mathbf{z}_{0}$ is an additive white Gaussian noise (AWGN) vector
whose elements have zero mean and variance $\sigma^{2}_{n}$.
Recall that $\mathbf{H}_{mn}$ represents the channel matrix from AP$_m$ to STA$_n$.
When $m=n$, we simply use one-letter subscript for conciseness.
$x_{0}$ and $x_{1}$ are data symbols sent from AP$_0$ and AP$_1$, respectively.
$E(\|x_{0}\|^{2})=E(\|x_{1}\|^{2})=1$ holds.
Since there is only one interference, data information carried by the steering
signal is $x^{s_t}_{k}=x_{1}$.
$P^{s_t}_{OH}$ denotes the power cost for IS at AP$_0$.
The variables $P^{s_t}_{k}$ and $\mathbf{p}^{s_t}_{k}$ are
the transmit power and precoder of the steering signal at AP$_k$.
$\mathbf{p}_{m}$ represents the precoding vector at AP$_m$ for data $x_{m}$.
In order to present our design under a fixed power constraint w.r.t. the victim user-pair,
we let AP$_0$ afford $P^{s_t}_{k}$, i.e., $P^{s_t}_{OH}=P^{s_t}_{k}$.
We adopt the singular value decomposition (SVD) based BF transmission, i.e.,
applying SVD to $\mathbf{H}_{m}$ to obtain $\mathbf{H}_{m}=\mathbf{U}_{m}\mathbf{D}_{m}
\mathbf{V}^{H}_{m}$ and employing $\mathbf{p}_{m}=\mathbf{v}^{(1)}_{m}$
where $\mathbf{v}^{(1)}_{m}$ represents the principal column vector of $\mathbf{V}_{m}$.
We can simply use $\mathbf{u}^{(1)}_{m}$ as the filter vector,
where $\mathbf{u}^{(1)}_{m}$ is the first column vector of $\mathbf{U}_{m}$.
Then, the estimated signal $\overline{y}_{m}=[\mathbf{u}^{(1)}_{m}]^{H}\mathbf{y}_{m}$.

%=======================================
% We delete this word on May 1st, 2016
%=======================================
%Since the interference can not only be mitigated but also be steered,
IS is designed as follows.
We first define the directions of desired signal and original interference combined
with steering signal as
$\mathbf{d}_{s}=\frac{\mathbf{H}_{0}\mathbf{p}_{0}}{\|\mathbf{H}_{0}\mathbf{p}_{0}\|}$ and
$\mathbf{d}_{i+s_{t,k}}=\frac{\sqrt{P_{T}}\mathbf{H}_{10}\mathbf{p}_{1}+\sqrt{P^{s_t}_{k}}
\mathbf{H}_{k0}\mathbf{p}^{s_t}_{k}}{\|\sqrt{P_{T}}\mathbf{H}_{10}\mathbf{p}_{1}
+\sqrt{P^{s_t}_{k}}\mathbf{H}_{k0}\mathbf{p}^{s_t}_{k}\|}$, respectively.
Then, by letting $\langle\mathbf{d}_{s},\mathbf{d}_{i+s_{t,k}}\rangle=0$, the original interference
can be steered to the orthogonal direction w.r.t.~the desired signal by the steering signal, $s_{t}$.

Since both the interference and the steering signal, denoted by $i$ and $s_{t,k}$, respectively,
can be decomposed into an in-phase component and a quadrature component,
denoted by superscripts $In$ and $Q$, respectively,
w.r.t.~$\mathbf{d}_{s}$, i.e., $i=i^{In}+i^{Q}$ and $s_{t,k}=s_{t,k}^{In}+s_{t,k}^{Q}$,
when $s_{t,k}^{In}=-i^{In}$, IS is realized.
%For simplicity, we omit the subscripts of interfering signal $i$ and steering signal $s_t$.
Fig.~3 illustrates the basic principle of IS.
A 2-dimensional representation is employed for readability.
The vectors in the figure indicate the spatial signals.
\begin{figure}[ht]
\vspace*{-0.1in}
%\begin{figure}[!htb]
%\graphicspath{/pdf}
\centering
\includegraphics[width=0.18\textwidth]{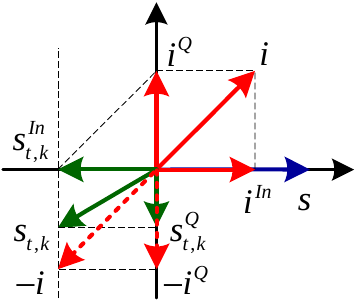}
\vspace{-5pt}
\caption{An illustration of interference steering.}
\label{fig: figure three}
%\end{figure}
\vspace*{-0.05in}
\end{figure}

It can be easily seen that given $s_{t,k}^{In}=-i^{In}$,
interference energy imposed on the desired transmission is countered.
As $s^{Q}_{t,k}$ approaches $-i^{Q}$, $s_{t,k}$ tends to $-i$,
i.e., IN is achieved.
Since the length of vector indicates the strength of signal, when $s^{Q}_{t,k}=0$,
we have $s_{t,k}=s^{In}_{t,k}$, and obtain an IS solution with minimum power overhead.

Based on the above discussion, we take the communication scenario depicted in Fig.~1
as an example. A general expression of IS implementation is then given as:
\begin{equation}
\left\{
\begin{array}{l}
\mathbf{p}^{s_t}_{k}=\mathbf{H}^{-1}_{k0}[s_{t,k}^{In}+s^{Q}_{t,k}]/\|\mathbf{H}^{-1}_{k0}
[s_{t,k}^{In}+s^{Q}_{t,k}]\|\\
P^{s_t}_{k}=P_{T}\|\mathbf{H}^{-1}_{k0}[s_{t,k}^{In}+s^{Q}_{t,k}]\|^2
\end{array}
\right.
\vspace*{-0.05in}
\end{equation}
%\\
where $s_{t,k}^{In}=-i^{In}=-\sqrt{P_{T}}\mathbf{P}\mathbf{H}_{10}\mathbf{p}_{1}$
and $\mathbf{P}=\mathbf{d}_{s}(\mathbf{d}^{T}_{s}\mathbf{d}_{s})^{-1}\mathbf{d}^{T}_{s}$
denotes projection matrix.
%============================================
%EE We delete this sentence on May 1st, 2016
%============================================
%As mentioned above,
When
$s_{t,k}^{Q}=-\sqrt{P_{T}}\mathbf{H}_{10}\mathbf{p}_{1}+\sqrt{P_{T}}\mathbf{P}
\mathbf{H}_{10}\mathbf{p}_{1}$,
we have $s_{t,k}=-i$ and IS becomes IN.
Then, Eq.~(2) can be rewritten as
\begin{equation}
\left\{
\begin{array}{l}
\mathbf{p}^{n_e}_{k}=-\mathbf{H}^{-1}_{k0}\mathbf{H}_{10}\mathbf{p}_{1}
/\|\mathbf{H}^{-1}_{k0}\mathbf{H}_{10}\mathbf{p}_{1}\|\\
P_{k}^{n_e}=P_{T}\|\mathbf{H}^{-1}_{k0}\mathbf{H}_{10}\mathbf{p}_{1}\|^2
\end{array}
\right..
%\vspace*{-0.05in}
\end{equation}

Note that the above results are obtained under $N_{t}=N_{r}$;
when $N_{t}>N_{r}$, the inverse of $\mathbf{H}_{k0}$ should be replaced by its Moore-Penrose
pseudo inverse.
%The mechanism can then be generalized.
One can now easily see that IS subsumes IN as a special case, making IS more general.
Considering the power overhead, we limit the sum of $P_{k}^{s_t}$ and power for the
desired signal's transmission to a fixed value $P_{T}$.
Moreover, only $s^{Q}_{t,k}=0$ is considered in our current design.
%=======================================
%EE We delete this word on May 1st, 2016
%=======================================
%In future, we will investigate the adaptation of $s^{Q}_{t,k}$ and its benefits.

Fig.~4 plots spatial spectrums of various signals to show the feasibility of IS
where $N_{t}=N_{r}=2$ and
the desired and interference signals are randomly generated.
The center frequency of input signal is $f_{0}=2.4$GHz, the antenna-element spacing
is half of the signal wavelength, and the SNR of each signal is $20$dB.
We first employ the MUSIC (MUltiple SIgnal Classification) algorithm to
estimate DoA (Direction of Arrival) of each signal.
Then, we reconstruct the spatial spectrum of signal components observed at the receiver.
For ease of comparison, we also plot signal transmission via the the second eigenmode
which is orthogonal to the desired signal utilizing the principal eigenmode.
\setcounter{figure}{3}
\begin{figure}[!htb]
%\vspace*{-0.05in}
%\graphicspath{/fig}
\centering
\includegraphics[width=0.34\textwidth,height=0.25\textwidth]{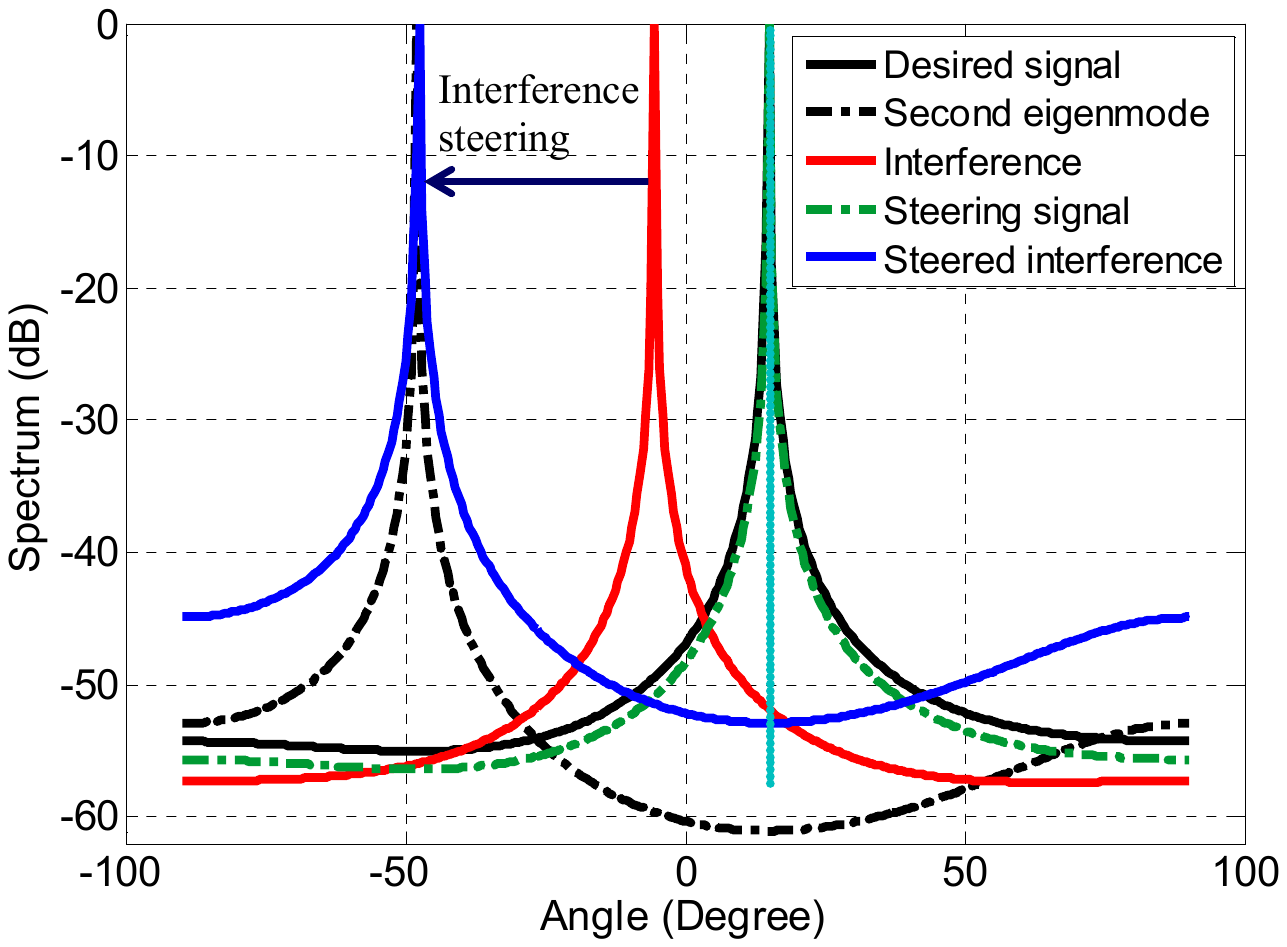}
\vspace*{-5pt}
\caption{Spatial spectrums of different signals.}
\label{fig: figure four}
\vspace*{-0.05in}
\end{figure}
%\begin{figure*}[!htb]
%\vspace*{-0.1in}
%\centering
%\begin{minipage}{1\columnwidth}
%\centering
%\includegraphics[width=0.85\columnwidth, height = 0.54\columnwidth]{fig/figure04.pdf}
%\vspace{-5pt}
%\caption{Spatial spectrums of different signals.}
%\label{fig: figure four}
%\end{minipage}
%\hfill
%\begin{minipage}{1\columnwidth}
%\centering
%\includegraphics[width=0.84\columnwidth, height = 0.54\columnwidth]{fig/figure05.pdf}
%\vspace{-5pt}
%\caption{Average system SE with different IM methods.}
%\label{fig: figure five}
%\end{minipage}
%\vspace*{-0.2in}
%\end{figure*}
%
%\begin{figure*}[!htb]
%\centering
%\begin{minipage}{1\columnwidth}
%\centering
%\includegraphics[width=0.7\columnwidth, height = 0.5\columnwidth]{fig/figure04.pdf}
%\vspace{-10pt}
%\caption{Spatial spectrum of different signals.}
%\label{fig: figure four}
%\end{minipage}
%\hfill
%\begin{minipage}{1\columnwidth}
%\centering
%\includegraphics[width=0.7\columnwidth, height = 0.5\columnwidth]{fig/figure05.pdf}
%\vspace{-10pt}
%\caption{Average system SE with different IM methods.}
%\label{fig: figure two}
%\end{minipage}
%\vspace*{-0.1in}
%\end{figure*}
Fig.~4 shows that the steering signal moves the interference's DoA to be the same as
the DoA of signal using the second eigenmode. Moreover, the lowest point of the steered
interference overlaps with the peak of the desired signal.
That is, the desired transmission has become interference-free.

\subsection{Comparison of IS and Other Transmitter-Based Methods}

So far, we have presented the basic signal processing of IS.
%CC The following description appears in the submitted manuscript,
%CC showing the differences between IS and other IM methods.
One may think IS similar to some existing IM methods, such as IA, ZFBF, CoMP, IN, etc., but it is not.
Basically, all of IA, ZFBF and CoMP require modifications at the interferer,
so that the transmission from an interfering Tx to its intended Rx is attenuated.
%FF Shin's edition, but I prefer the original one.
%so as to attenuate the transmission from an interfering Tx to its intended Rx.
In case of CoMP, the user is served by multiple APs cooperatively.
IN and IS are victim Tx-side implementations which are suitable only when
the interferer is willing to sacrifice as is usually the case.
%FF Shin's edition, sacrifice is more precise.
%reduce its transmit power.
Compared to IN, IS focuses on canceling only the effective part of the interference,
thus becoming more power-efficient. However, it costs one DoF at Rx, just as IA.

Without specifications, the following simulation results in this paper are under $N_{t}=N_{r}=2$.
However, the same conclusion can be drawn with various antenna settings.
For space limitation, we omit the results for other $N_{t}$ and $N_{r}$ values.
Fig.~5 comparatively evaluates the achievable system SE of IA, ZFBF, IN and IS by
%CC To indicate how we obtain Fig. 5.
using MATLAB simulation and
with the consideration of two communication pairs ($K=2$) --- from AP$_{0}$ to STA$_0$ and
from AP$_{1}$ to STA$_1$.
AP$_1$ interferes with STA$_0$.
Note that IA and ZFBF are implemented at AP$_1$, incurring performance loss to
the transmission from AP$_1$ to STA$_1$.
As for ZFBF, since $N_{t}$ should be no less than the total number of receiving antennas
across all receivers, we let $N_{r}=1$. With the other three methods, $N_{t}=N_{r}=2$.
IN and IS are carried out by AP$_0$, when their power overheads, $P^{n_e}_{0}$ and
$P^{s_t}_{0}$, exceed the power budget of an AP, e.g., $P_{T}$,
we simply regard the victim transmission's SE as 0.
In the figure, SE of an interference-free point-to-point MIMO (p2pMIMO)
transmission employing BF is also plotted as a reference.
IS is shown to yield the best system SE (SE of $K=2$ transmissions).
Since $N_r=1$, the SE performance of ZFBF is inferior to that of IS and IA.
System SE of IN is close to that of p2pMIMO,
so SE of the transmission pair using IN is inferred to be very low.
It should be noted that due to the randomness of wireless channels,
IS may incur more cost compared to IA,
thus outputting lower SE under some channel conditions.
However, as shown in the figure,
the system SE of IS is statistically higher than that of IA.
\setcounter{figure}{4}
\begin{figure}[!htb]
\vspace*{-0.05in}
%\graphicspath{/fig}
\centering
\includegraphics[width=0.34\textwidth,height=0.25\textwidth]{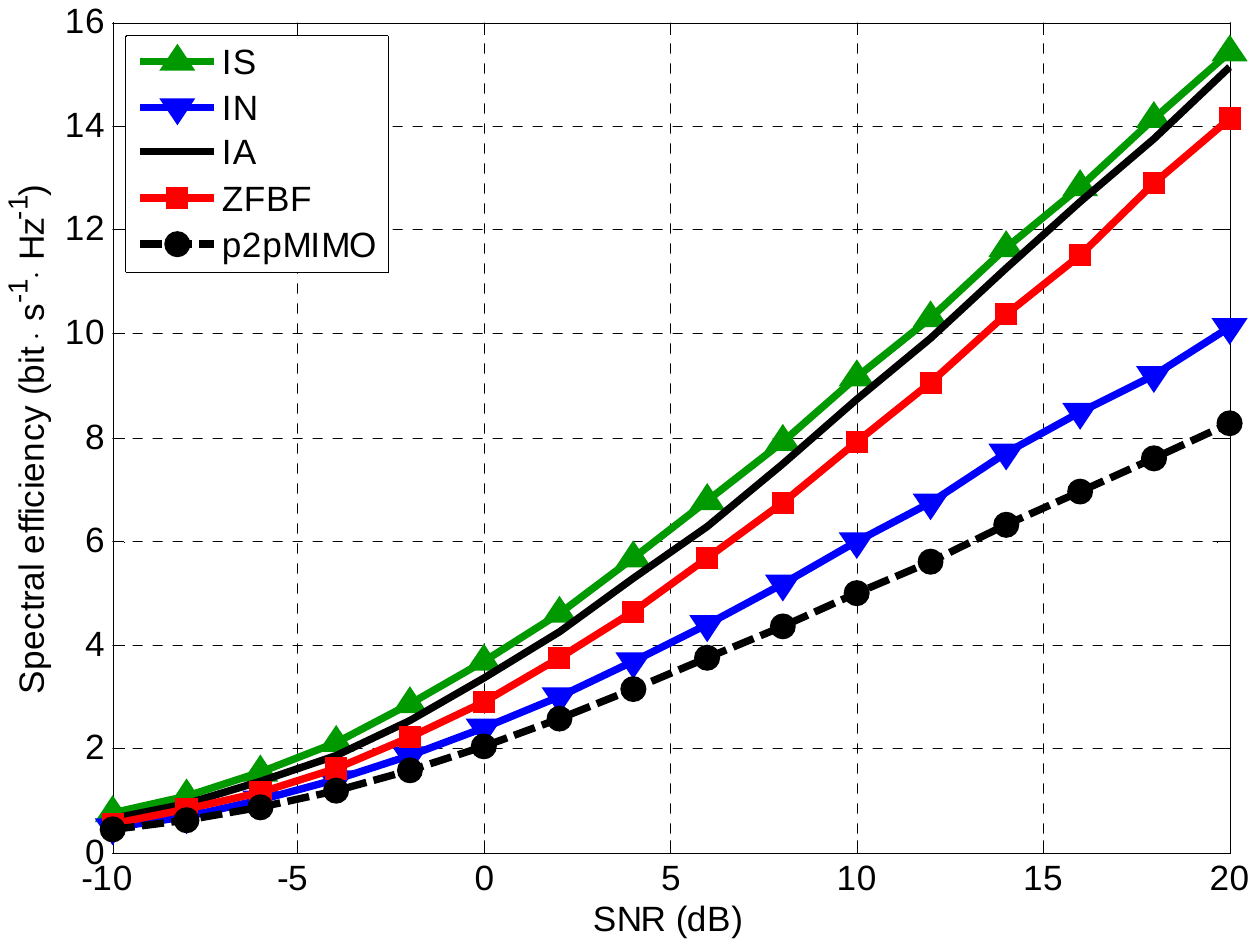}
\vspace*{-5pt}
\caption{Average system SE with different IM methods.}
\label{fig: figure five}
\vspace*{-0.05in}
\end{figure}

\subsection{Multi-Interference Processing}

In the above discussion, only one interferer is considered.
When a victim Rx suffers from multiple interferers,
its associated AP asks the WLAN controller for assistance.
The latter is able to check these interference components, combine them together,
and then calculate an IS solution in terms of this combined interference
%CC Add the infocom decoding paper.
[21].
In other words, multiple independent interferences are not treated individually but as a whole
(after combining them), and hence the dimension of interference is reduced to 1.
In the case of multi-interferences, Eq.~(1) can be rewritten as
\vspace*{-0.05in}
\begin{equation}
\begin{aligned}
\mathbf{y}_{0}&=\sqrt{P_{T}-P^{s_t}_{k}}\mathbf{H}_{0}\mathbf{p}_{0}x_{0}+\sum_{m\in\{1,\cdots,M\}}\sqrt{{P}_{T}}\mathbf{H}_{m0}\mathbf{p}_{m}x_{m}\\
&+\sqrt{P^{s_t}_{k}}\mathbf{H}_{k0}\mathbf{p}^{s_t}_{k}x^{s_t}_{k}+\mathbf{z}_{0}
\end{aligned}
\vspace*{-0.05in}
\end{equation}
\noindent
where $m$ and $M$ are the index and the total number of interferences, respectively.
Transmission from AP$_0$ to STA$_0$ is interfered with by the other $M=K-1$ transmissions.
$x^{s_t}_{k}$ is dependent on the combination of all interferences.
%\begin{figure}[!htb]
%\graphicspath{/fig}
%\centering
%\includegraphics[width=0.36\textwidth,height=0.27\textwidth]{fig/figure062.pdf}
%\caption{Average system spectral efficiency under $M\in\{1,2,3,4\}$ and SNR=$10$dB.}
%\label{fig: figure six}
%\vspace*{-0.1in}
%\end{figure}
\setcounter{figure}{5}
\begin{figure*}[!htb]
\vspace*{-0.1in}
\centering
\begin{minipage}{0.66\columnwidth}
\centering
\includegraphics[width=1.01\columnwidth, height = 1.72in]{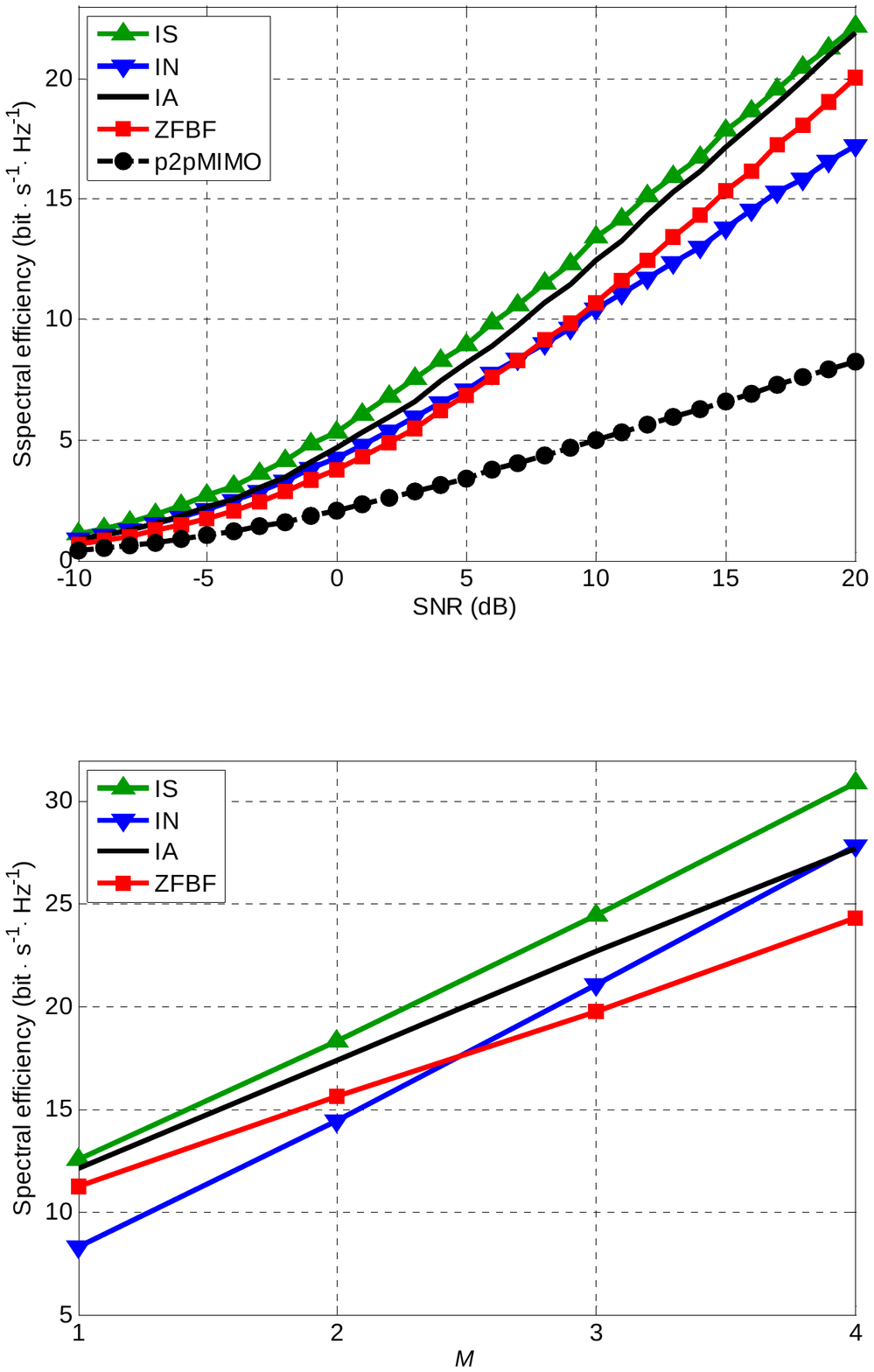}
\vspace{-14pt}
\caption{Average system SE under $M\in\{1,2,3,4\}$ and SNR=$15$dB.}
\label{fig: figure six}
\end{minipage}
\hfill
\begin{minipage}{0.66\columnwidth}
\centering
\includegraphics[width=1.01\columnwidth, height = 1.73in]{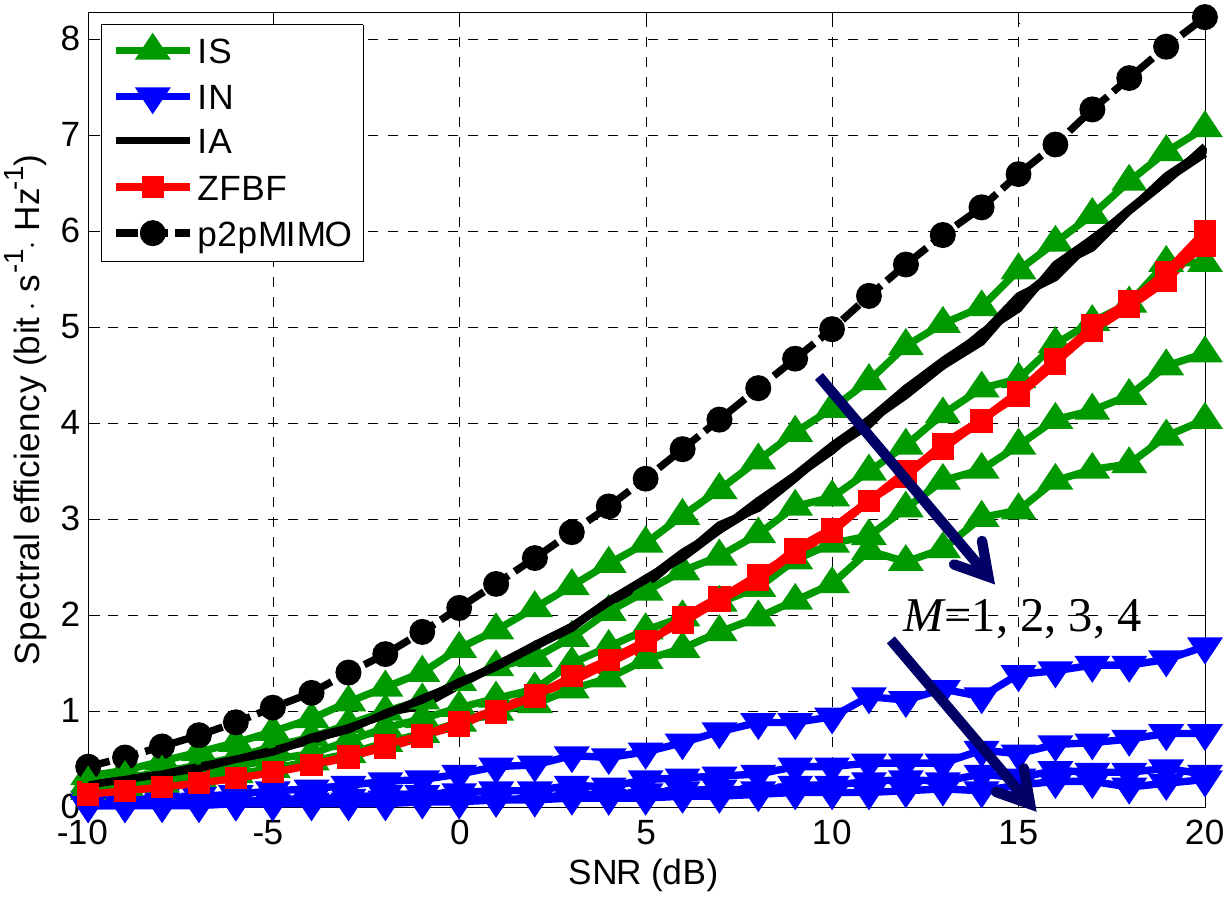}
\vspace{-14pt}
\caption{SE of a single transmission pair under $M\in\{1,2,3,4\}$.}
\label{fig: figure seven}
\end{minipage}
\hfill
\begin{minipage}{0.66\columnwidth}
\centering
\includegraphics[width=1.01\columnwidth, height = 1.74in]{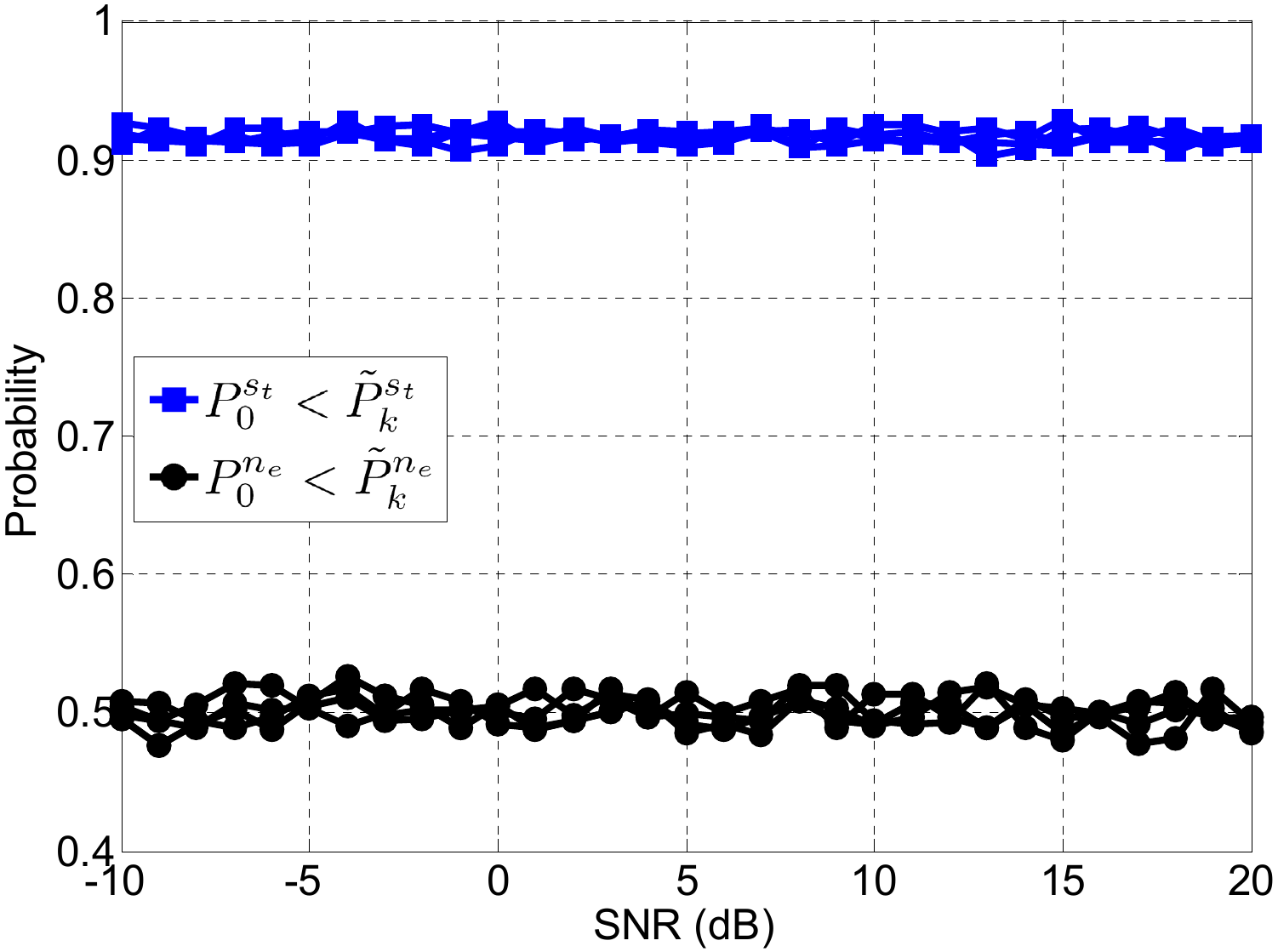}
\vspace{-14pt}
\caption{Comparison of power overheads under $M\in\{1,2,3,4\}$.}
\label{fig: figure eight}
\end{minipage}
\vspace*{-0.1in}
\end{figure*}

%CC To indicate how we obtain Fig. 6 and Fig. 7.
As in Fig.~6, MATLAB simulation is employed
to evaluate the overall SE of $K=M+1$ transmissions
of which $M$ are interferers and 1 is the victim, for different IM methods.
For simplicity, we assume no interference between any two interferers.
As shown in Fig.~6, the system SE improves as $M$ increases.
With IA and ZFBF, the interfering Tx conducts IM, thus degrading all
transmissions from the interfering Txs to their intended Rxs.
In case of IS and IN, AP$_0$ performs the IM, only the transmission to the the victim is affected.
However, IS incurs much less power cost than IN, and
hence yields higher SE than IN under a transmit power constraint.
Thus, IS is shown to provide the best performance, while IN's SE gradually
approaches IA's and exceeds ZFBF's as $M$ increases.

We use $r_{s_{0}}^{\mathcal{M}}$, $r_{s_m}^{\mathcal{M}}$ and $R^{\mathcal{M}}_{sys}$ to denote
the average SE of the victim AP$_0$'s transmission to STA$_0$, the
interferer's own transmission from AP$_m$ to STA$_m$ where $m\in\{1,\cdots,K-1\}$
and the system employing $\mathcal{M}$ as the IM method, respectively.
$r_{BF}$ is the SE of p2pMIMO with BF.
Without loss of generality, we let the victim AP$_0$ carry out IN or IS.
Then, the average system SE of IS, IN, IA and ZFBF with $M$ interferers can be calculated as
$R^{IS}_{sys}=r_{s_{0}}^{IS}+Mr_{BF}$, $R^{IN}_{sys}=r_{s_{0}}^{IN}+Mr_{BF}$,
$R^{IA}_{sys}=r_{BF}+\sum^{M}_{m}r_{s_m}^{IA}$, $R^{ZFBF}_{sys}=r_{BF}
+\sum^{M}_{m}r_{s_m}^{ZFBF}$.
For example, when IS or IN is employed, the SE of the victim's transmission is
$r_{s_{0}}^{IS}$ or $r_{s_{0}}^{IN}$, whereas for the other transmissions, SE is calculated
in terms of p2pMIMO.
That is, although $r_{s_{0}}^{IS}$ and $r_{s_{0}}^{IN}$ decreases as $M$ grows,
$R^{IS}_{sys}$ and $R^{IN}_{sys}$ are dominated by $Mr_{BF}$ which grows linearly
with $M$, thus enhancing the system SE.
When IN or IS is implemented at AP$_k$ where $k\in\{1,\cdots,M\}$, the AP$_0$'s
transmission to STA$_0$ is free of interference, and hence its SE is equal to $r_{BF}$.
We then have $R^{IS}_{sys}=r_{s_{k}}^{IS}+Mr_{BF}$ and
$R^{IN}_{sys}=r_{s_{k}}^{IN}+Mr_{BF}$.
%The above analysis is directly applied.
In case of IA and ZFBF, the victim's transmission becomes free of interference,
and hence its SE is equal to $r_{BF}$.
Since $r_{s_m}^{IA}<r_{BF}$ and $r_{s_m}^{ZFBF}<r_{BF}$,
protection of STA$_0$ from the interference of AP$_m$ ($m\in\{1,\cdots,M\}$)
requires the sacrifice of SE of all the interfering transmissions.
%\begin{figure}[!htb]
%\graphicspath{/fig}
%\centering
%\includegraphics[width=0.36\textwidth,height=0.27\textwidth]{fig/figure063.pdf}
%\caption{Single transmission pair spectral efficiency under $M\in\{1,2,3,4\}$.}
%\label{fig: figure seven}
%\vspace*{-0.1in}
%\end{figure}

Fig.~7 shows the SE of a single transmission pair for different IM methods.
AP$_0$'s transmission to STA$_0$ is interfered with by the other $M=K-1$ transmissions.
Under IS and IN, we let AP$_0$ perform IM, and plot $r_{s_{0}}^{IS}$ and $r_{s_{0}}^{IN}$,
while studying the SE of an arbitrary interfering transmission --- i.e., $r_{s_{m}}^{IA}$ and
$r_{s_{m}}^{ZFBF}$, $m\in\{1,\cdots,M\}$ --- under IA and ZFBF.
Both $r_{s_{0}}^{IS}$ and $r_{s_{0}}^{IN}$ are shown to decrease as $M$ increases.
In case of the transmission pair with IA or ZFBF, there is no interference to it,
thus making its SE independent of $M$.
From Fig.~7 we can obtain the SE loss of one communication pair which implements an IM method
by subtracting the SE of the corresponding method from that of p2pMIMO.
With IS and IN, the victim's transmission suffers from some SE loss,
whereas for ZFBF and IA, each interferer sacrifices its SE performance so as to avoid interference
to the victim Rx. As a result, for IS and IN, in order to guarantee the victim's transmission performance,
the number of interferences to a Rx, denoted by $\eta$ (see in Section V
%=======================================
%EE We delete this word on May 1st, 2016
%=======================================
%for details
), should be limited.

\subsection{No Need for Cooperative IS}

%We now consider where IS should be implemented.
Two types of IS are conceivable: cooperative and non-cooperative.
In case of the former, IS is performed by the interfering AP or the other adjacent
AP that the victim STA is not associated with, whereas
the victim AP generates the IS signal in case of the latter.
Note, however, both schemes require the collaboration, i.e., CSI and data sharing, between
the AP performing IS and the interfering AP.
We elaborate below why cooperative IS is not necessary from two aspects.

First, there may not exist strong motivation for the interfering AP to implement IS,
since such an implementation will deteriorate its own transmission and there is not enough
incentive to help the other AP's client.
%there doesn't exist any motivation for the interfering AP to implement IS,
%since such an implementation will deteriorate its own transmission and there is no
%incentive to help the other AP's client.
Take the communication scenario depicted in Fig.~1 as an example.
If AP$_1$ carries out IS for STA$_0$, the received signal at STA$_1$ becomes
\begin{equation}
\vspace*{-0.05in}
\mathbf{y}_{1}=\sqrt{P_{T}-P^{s_t}_{1}}\mathbf{H}_{1}\mathbf{p}_{1}x_{1}+\sqrt{P^{s_t}_{1}}
\mathbf{H}_{1}\mathbf{p}^{s_t}_{1}x_{1}^{s_t}+\mathbf{z}_{1}
\vspace*{-0.05in}
\end{equation}
where $x_{1}^{s_t}=x_{1}$. By employing $\mathbf{u}^{(1)}_{1}$ as the receive filter,
and noting that the first two signal components on the RHS of Eq.~(5)
are neither in the same direction nor orthogonal,
the received SNR of STA$_1$ satisfies the following inequality:
%\begin{equation}
%\vspace*{-0.03in}
%\gamma_{1}<\frac{1}{\sigma_{n}^{2}}\left\{
%(P_{T}-P^{s_t}_1)[\lambda^{(1)}_{1}]^{2}+P^{s_t}_{1}\|[\mathbf{u}^{(1)}_{1}]^{H}
%\mathbf{H}_{1}\mathbf{p}^{s_t}_{1}]\|^{2}
%\right\}.
%%\vspace*{-0.03in}
%\end{equation}
\begin{equation}
\vspace*{-0.03in}
\gamma_{1}<\left\{
(P_{T}-P^{s_t}_1)[\lambda^{(1)}_{1}]^{2}+P^{s_t}_{1}\|[\mathbf{u}^{(1)}_{1}]^{H}
\mathbf{H}_{1}\mathbf{p}^{s_t}_{1}]\|^{2}
\right\}/{\sigma_{n}^{2}}.
%\vspace*{-0.03in}
\end{equation}
$\lambda^{(1)}_{1}$ represents the principal eigenvalue of $\mathbf{H}_{1}$.
Since $\|[\mathbf{u}^{(1)}_{1}]^{H}\mathbf{H}_{1}\mathbf{p}^{s_t}_{1}]\|<\lambda^{(1)}_{1}$,
we can have
%$\gamma_{1}<\frac{P_{T}[\lambda^{(1)}_{1}]^2}{\sigma_{n}^{2}}$
$\gamma_{1}<{P_{T}[\lambda^{(1)}_{1}]^2}/{\sigma_{n}^{2}}$
where
%$\frac{P_{T}[\lambda^{(1)}_{1}]^2}{\sigma_{n}^{2}}$
${P_{T}[\lambda^{(1)}_{1}]^2}/{\sigma_{n}^{2}}$ is the received SNR of STA$_1$
without implementing IS. That is, there will be some SE loss w.r.t.~the transmission in BSS$_1$.

Then, let's consider the performance of IS implemented by the victim STA's nearby AP
other than its associated AP and the interfering AP.
In this case, for ease of comparison we let AP$_{0}$ be responsible for the power overhead of IS,
and the sum of $P_{k}^{s_t}$ and power for the
desired signal's transmission is limited to $P_{T}$.
According to Fig.~1 and the design
%=======================================
%EE We delete this word on May 1st, 2016
%=======================================
%described
in Section III.A,
the received SNR of STA$_0$ after post-processing is
%\vspace*{-0.1in}
%\begin{equation}
%\gamma_{0}=\frac{1}{\sigma_{n}^{2}}(P_{T}-P^{s_t}_k)[\lambda^{(1)}_{0}]^{2}.
%\vspace*{-0.1in}
%\end{equation}
\vspace*{-0.05in}
\begin{equation}
\gamma_{0}=(P_{T}-P^{s_t}_k)[\lambda^{(1)}_{0}]^{2}/{\sigma_{n}^{2}}.
\vspace*{-0.05in}
\end{equation}

It can be easily seen from Eq.~(7) that $\gamma_{0}$ is determined by the power cost of IS.
If IS is carried out by AP$_0$, $P^{s_t}_k$ is given by $P^{s_t}_{0}=P_{T}\|\mathbf{H}^{-1}_{0}
\mathbf{P}\mathbf{H}_{10}\mathbf{p}_{1}\|^2$.
Substituting $\mathbf{P}=\mathbf{d}_{s}(\mathbf{d}^{T}_{s}\mathbf{d}_{s})^{-1}\mathbf{d}^{T}_{s}$
into $P^{s_t}_{0}$, we can obtain the power overhead of non-cooperative IS as:
%\vspace*{-0.1in}
%\begin{equation}
%P^{s_t}_{0}=\frac{P_{T}}{\|\mathbf{H}_{0}\mathbf{p}_{0}\|}\|\mathbf{H}^{-1}_{0}\mathbf{H}_{0}\mathbf{p}_{0}\|^{2}\psi
%=\frac{P_{T}}{\|\mathbf{H}_{0}\mathbf{p}_{0}\|}\psi
%\vspace*{-0.05in}
%\end{equation}
\vspace*{-0.05in}
\begin{equation}
P^{s_t}_{0}={P_{T}}{\|\mathbf{H}_{0}\mathbf{p}_{0}\|}^{-1}\|\mathbf{H}^{-1}_{0}\mathbf{H}_{0}\mathbf{p}_{0}\|^{2}\psi
={P_{T}}{\|\mathbf{H}_{0}\mathbf{p}_{0}\|}^{-1}\psi
\vspace*{-0.05in}
\end{equation}
where $\psi=\|(\mathbf{d}^{T}_{s}\mathbf{d}_{s})^{-1}\mathbf{d}^{T}_{s}\mathbf{H}_{10}\mathbf{p}_{1}\|^{2}$.
Similarly, when AP$_k$ implements IS for STA$_0$ where $k\not\in\{0,1\}$,
the power consumption of cooperative IS is
%\vspace*{-0.05in}
%\begin{equation}
%\tilde{P}^{s_t}_{0}=\frac{P_{T}}{\|\mathbf{H}_{0}\mathbf{p}_{0}\|}\|\mathbf{H}^{-1}_{k0}
%\mathbf{H}_{0}\mathbf{p}_{0}\|^{2}\psi.
%\vspace*{-0.05in}
%\end{equation}
\vspace*{-0.05in}
\begin{equation}
\tilde{P}^{s_t}_{k}={P_{T}}{\|\mathbf{H}_{0}\mathbf{p}_{0}\|}^{-1}\|\mathbf{H}^{-1}_{k0}
\mathbf{H}_{0}\mathbf{p}_{0}\|^{2}\psi.
\vspace*{-0.05in}
\end{equation}

%\begin{figure}[!htb]
%\graphicspath{/fig}
%\centering
%\includegraphics[width=0.36\textwidth,height=0.27\textwidth]{fig/figure08.pdf}
%\vspace{-10pt}
%\caption{Power overhead comparison of non-cooperative IS/IN and cooperative IS/IN under $M\in\{1,2,3,4\}$.}
%\label{fig: figure eight}
%\vspace*{-0.05in}
%\end{figure}
We use MATLAB simulation to obtain the probability that $P^{s_t}_{0}$ is less than
$\tilde{P}^{s_t}_{k}$, or equivalently $\|\mathbf{H}^{-1}_{k0}\mathbf{H}_{0}\mathbf{p}_{0}\|<1$,
as shown in Fig.~8.
Although the above derivation is based on the single-interference assumption,
the results for multiple interferences can also be derived, as shown in the figure.
Since IN is a special case of IS, the power overheads of non-cooperative and cooperative IN,
denoted by $P^{n_e}_{0}$ and $\tilde{P}^{n_e}_{k}$ respectively,
are also studied.
The probability of $P^{s_t}_{0}<\tilde{P}^{s_t}_{k}$ is notably higher than that of
$P^{n_e}_{0}<\tilde{P}^{n_e}_{k}$, and both are independent of $M$.
This is because cooperative IS consumes more power than the non-cooperative one,
whereas for IN, the power cost is statistically the same with or without collaboration.
So, we conclude that IS should better be implemented at the victim transmitter,
whereas for IN, selective diversity gain can be obtained
by adaptively choosing candidate Txs which are willing to assist the victim Rx in IN.

\section{Application of IS in Enterprise WLANs}

In this section, we elaborate the application of IS in an enterprise WLAN.
Fig.~9 shows an example network with $K=6$ BSSs.
However, our discussion can be extended to a general network
with a random topology and arbitrary number of interferences.
Since an AP can only serve one STA on a frequency channel at a time,
other users with transmission demand will be moved to different channels.
Without loss of generality, we use STA$_m$ ($m\in\{0,\cdots,K-1\}$)
to represent the current client served by AP$_m$.
Since there are overlapping areas among adjacent BSSs and
all BSSs reuse the same frequency channel, CCI should be well managed so that
multiple interfering transmissions can be supported simultaneously.
In the next two subsections, we will first discuss the maintenance of information required
for IS and use graph theory to analyze the interference relationship among multiple BSSs.
We will then elaborate the realization of IS at different network entities.
\setcounter{figure}{8}
\begin{figure}[ht]
%\vspace*{-0.05in}
%\graphicspath{/fig}
\centering
\includegraphics[width=0.43\textwidth]{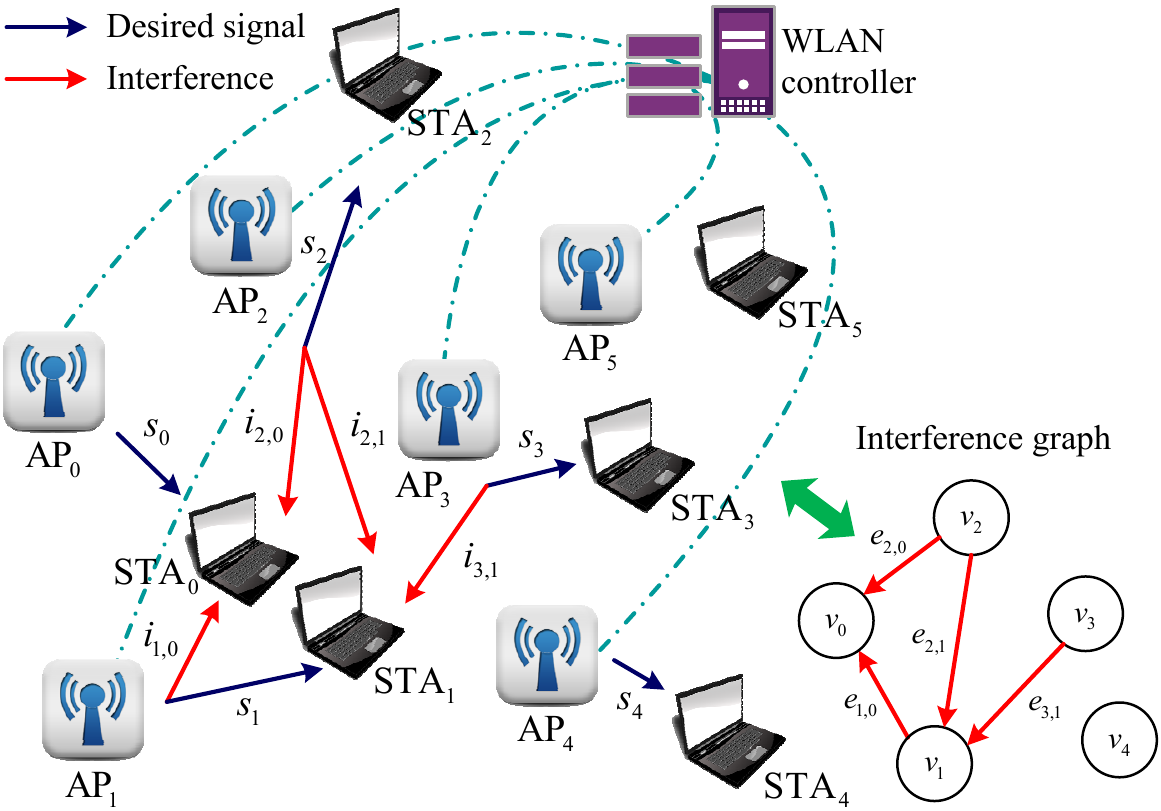}
\vspace{-5pt}
\caption{An example WLAN with multiple interferences and its interference graph.}
\label{fig: figure nine}
%\vspace*{-0.1in}
\end{figure}

\subsection{Data Structure and Graph Analysis}

In order to achieve IS, a proper data structure should be used for
managing necessary network information.
We adopt a {\em connection matrix} $\mathbf{C}$ and a {\em transmitting status
vector} $\mathbf{T}$ for this purpose.
$\mathbf{C}$ and $\mathbf{T}$ of the WLAN depicted in Fig.~9 are then expressed as
\vspace*{-0.05in}
\begin{equation}
\mathbf{C}=
\left[
\begin{array}{cccccc}
1 & 1 & 1 & 0 & 0 & 0 \\
0 & 1 & 1 & 1 & 0 & 0 \\
0 & 0 & 1 & 0 & 0 & 0 \\
0 & 0 & 0 & 1 & 0 & 1 \\
0 & 0 & 0 & 0 & 1 & 0 \\
0 & 0 & 0 & 0 & 0 & 1 \\
\end{array}
\right]
\vspace*{-0.1in}
\end{equation}
\\
and
\\
\vspace*{-0.1in}
\begin{equation}
\mathbf{T}=\left[
1~1~1~1~1~0
\right].
\vspace*{-0.05in}
\end{equation}

As shown in Eq.~(10), if STA$_n$ can (not) hear AP$_m$, then its connection status
$\mathbf{C}(n+1,m+1) =  1(0)$.
The  principal diagonal elements of $\mathbf{C}$ are 1 due to the association
between the corresponding APs and STAs.
For example, STA$_0$ is associated with AP$_0$, and hence $\mathbf{C}(1,1)=1$.
STA$_0$ can also hear adjacent non-associated APs, say AP$_1$ and AP$_2$,
then $\mathbf{C}(1,2)=\mathbf{C}(1,3)=1$.
In case of STA$_3$, although AP$_5$ is not transmitting to STA$_5$,
STA$_3$ is in the coverage of AP$_5$, and thus $\mathbf{C}(4,5)=1$.
From Eq.~(11) we can see that when AP$_m$ transmits to its associated client STA$_m$,
the $(m+1)^{th}$ element of $\mathbf{T}$, $\mathbf{T}(m)$, is 1.

Both $\mathbf{C}$ and $\mathbf{T}$ are constructed based on the feedbacks from all APs
and maintained by the WLAN controller.
When a collision is detected and reported to an AP,
the AP will inquire the central controller.
Upon receiving the interference management requirement,
the controller performs the operation in Eq.~(12) to obtain
the {\em adjacency matrix}, or the {\em interference matrix},
represented by $\mathbf{A}$ as:
\vspace*{-0.07in}
\begin{equation}
\mathbf{A}=\textrm{diag}(\mathbf{T})(\mathbf{C}-\mathbf{I})
\vspace*{-0.07in}
\end{equation}
where $\textrm{diag}(\mathbf{a})$ indicates the diagonalization of vector $\mathbf{a}$
and $\mathbf{I}$ is a $K\times K$ unit matrix.
$\mathbf{A}$ specifies the interferences among all BSSs.
For example, substituting Eqs.~(10) and (11) into (12),
we can get
\vspace*{-0.07in}
\begin{equation}
\mathbf{A}=
\left[
\begin{array}{cccccc}
0 & 1 & 1 & 0 & 0 & 0 \\
0 & 0 & 1 & 1 & 0 & 0 \\
0 & 0 & 0 & 0 & 0 & 0 \\
0 & 0 & 0 & 0 & 0 & 1 \\
0 & 0 & 0 & 0 & 0 & 0 \\
0 & 0 & 0 & 0 & 0 & 0 \\
\end{array}
\right],
\vspace*{-0.05in}
\end{equation}
based on which an IS solution for the network can be calculated as
discussed in the next subsection.

So far, the network's interference status is represented by $\mathbf{A}$.
According to graph theory, an interference graph can also be employed to
describe the interference relationship among BSSs. As shown in the bottom-right of Fig.~9,
%\textcolor{blue}{
%which is a concise expression of Fig.~9,}
each vertex denoted by $v_m$ indicates a data transmission in a BSS.
The directed edge $e_{m,n}$ represents the interference from vertex $m$ to $n$.
The weight of an edge could be the strength of interference to its destination.
Since AP$_5$ is not in service, $v_5$ is not included in the graph.
%=======================================================
%EE We combine this figure with Fig. 9 on May 1st, 2016
%=======================================================
%\begin{figure}[ht]
%\vspace*{-0.1in}
%\graphicspath{/fig}
%\centering
%\includegraphics[width=0.17\textwidth]{fig/figure10.pdf}
%\vspace{-5pt}
%\caption{An example interference graph.}
%\label{fig: figure ten}
%\vspace*{-0.2in}
%\end{figure}

Before delving into interference further, we first discuss some useful
properties of the interference graph.

First, since we assume use of BF by all transmissions,
the number of edges between two adjacent vertices is at most one.
However, our design can be easily extended to the case where
spatial multiplexing (SM) is used by the interferer and/or the victim BSS.
Such an extension does not affect the applicability of IS.
On one hand, when the interferer adopts SM, there exist multiple interference components,
but since IS is designed based on the aggregated effect of interference, IS is achievable
in the same way as in the single-interference case.
On the other hand, when the victim BSS employs SM, there will be multiple mutually
orthogonal intended transmissions from the victim AP to its recipient.
Since each steering signal is in the opposite direction to the corresponding desired signal
component, the steering signal for one spatial data stream will not interfere with the others
and an IS solution for the STA receiving multiple desired streams is available.

Second, an AP cannot implement IS for more than one victim STA simultaneously
due to the non-orthogonality of transmissions to these STAs.
For example, if an AP serves two clients both of which are interfered with by nearby AP(s),
when the victim AP generates a steering signal for one victim user,
additional interference (regarded as the side-effect of IS) will be incurred to the other victim
transmission, and vice versa, thus making IS unavailable.

Third, we don't advocate cooperative IS due to its higher power overhead
than the non-cooperative counterpart.
In addition, an AP's assistance of another AP's client in achieving IS
will incur additional interference to ongoing transmissions of
its associated STA as well as non-associated STAs in its coverage area.
An IS solution that counts for assistant AP selection will be very difficult, or even
impossible to obtain due to the increased computational complexity or
because the problem may become non-convergent.
For example, when AP$_k$ is employed for sending a steering signal $s_{t,k}$ to
assist AP$_0$ in IS, as shown in Fig.~1,
an additional interference $i'_{k,1}$ will be imposed on STA$_1$.
If AP$_1$ adjusts its transmission to adapt to this interference,
the interference from AP$_1$ to STA$_0$, i.e., $i_{1,0}$ varies, needing an updated
steering signal from AP$_k$. Since the above process is non-convergent,
cooperative IS is unavailable in such a situation.

Fourth, recall that for the first case in Fig.~2, an interference cycle,
i.e., mutual-interference between two BSSs, occurs, and
then one transmission is blocked.
When this situation is generalized to the multi-BSS ($K\geq3$) scenario,
similarly to the two-BSS case, we disallow cycles, i.e., if there is (are) cycle(s),
then at least one vertex should be deleted so as to break cycle(s).
This can be explained in terms of the side effects of IS, i.e., additional interference.
Take a three-node cycle as an example in Fig.~9
where $e_{2,0}$ is replaced with $e_{0,2}$, forming a cycle with
$v_0$, $e_{0,2}$, $v_2$, $e_{2,1}$, $v_1$ and $e_{1,0}$.
Without loss of generality, we calculate an IS solution for $v_{0}$ first and
then obtain $s_{t,0}$. Since $s_{t,0}$ introduces additional interference $i'_{0,2}$
to $v_{2}$, a steering signal for $v_{2}$ should be generated based on $i_{0,2}+i'_{0,2}$.
Similarly, $s_{t,1}$ is calculated in terms of $i_{2,1}+i_{0,2}+i'_{0,2}$.
In the end, $i'_{1,0}$ yielded by $s_{t,1}$ will lead to the recalculation of $s_{t,0}$ at $v_{0}$.
This phenomenon is similar to a positive feedback that makes
a stable IS solution unavailable in a network with interference cycles.

Based on the above discussion, the feasibility characterization of IS can be
stated as in the following theorem.

\textbf{Theorem 1}: A set of directed links are feasible for IS if and only if their interference graph is acyclic.

To be specific, in order to achieve IS, the original interference graph
should be converted to a directed acyclic graph (DAG) [22].
Moreover, due to the randomness of networks, the interference graph may not be
connected. Thus, given an interference graph, we should first check its connectivity.
For each connected part, we employ a depth-first search (DFS) algorithm to detect
cycles [22]. Considering the fact that each vertex represents a data transmission,
we break a cycle by deleting a set of nodes according to the following two rules.
First, remove as few vertices as possible. Second, if the first rule is met,
the sum weight of the outgoing edges of the deleted node set should be maximized,
i.e., remove the set of nodes generating the most interference.
A brute-force search can be used to obtain such a set of nodes.

So far, we can obtain one or multiple DAGs, depending on the connectivity of the
original interference graph. Then, an IS solution is calculated for each sub-graph.
Topological sorting [22] is used to determine the order of vertices that IS is computed for.
Before detailing the application of this algorithm in the next subsection,
we introduce one definition and two properties of the DAG of our interest as follows.

\textbf{Definition 1}: For a vertex, the number of tail ends adjacent to a vertex is called
the {\em indegree} of the vertex and the number of head ends adjacent to a vertex is its {\em outdegree}.

\textbf{Property 1}: Every acyclic graph contains at least one vertex with zero indegree;
otherwise, there is a cycle.

\textit{Proof}: The proof can be found in [23]. $\blacksquare$

\textbf{Property 2}: The interference graph remains unchanged with non-cooperative IS.

\textit{Proof}: Note that each vertex represents a data transmission and
the transmission range of a steering signal is the same as its data signal.
Given an interference graph, if there is an edge, say $e_{m,n}$, between two vertices,
and AP$_m$ generates steering signal $s_{t,m}$ for its client, then the interference $i'_{m,n}$
caused by $s_{t,m}$ in addition to the original interference $i_{m,n}$
will be incurred to STA$_n$ being served by AP$_n$;
if there is no edge between $v_{m}$ and $v_n$,
after AP$_m$ implements IS, STA$_n$ receives nothing from AP$_m$.
For the above two cases, no new edge between the two nodes will be established
with non-cooperative IS, and thus Property 2 follows. $\blacksquare$

\subsection{Application of IS in WLANs}

We now detail the implementation of IS at STA, AP, and the central controller, respectively.
Note that the interferer is not required to do anything for IS except for CSI and data sharing with the victim transmitter.

A STA  needs to perform 6 tasks for IS as follows. First, it must construct a list of available
networks in terms of the received beacon signals.
Second, it must determine whether it is an edge STA or a central STA
according to the number of available networks on the list.
Third, it should estimate CSI based on the NDP frames sent from nearby APs
and feed back this information to the associated AP.
Fourth, it should contend for a channel according to the LBT rule.
Fifth, it should detect collision/interference from adjacent APs during data transmission.
Sixth, it should ask the associated AP for assistance when collision/interference occurs.

It should be noted that only edge STA requests for IM assistance,
i.e., the proposed mechanism focuses on the management of interference in the overlapping areas.
Compared to the existing IEEE 802.11 protocols,
one can see that IS requires only minor modifications of STA,
i.e., an increased CSI estimation workload and feedback overhead w.r.t. the edge STA.

APs are connected to a central controller, and only need to inquire of the controller
upon reception of an IM request from their clients. The controller calculates IS.
In future WLANs, more processing is expected to move from APs to the controller,
i.e., an AP may only be responsible for generating RF signals
by following the instructions sent from the controller.
\setcounter{figure}{9}
\begin{figure}[ht]
%\vspace*{-0.1in}
%\graphicspath{/fig}
\centering
\includegraphics[width=0.47\textwidth]{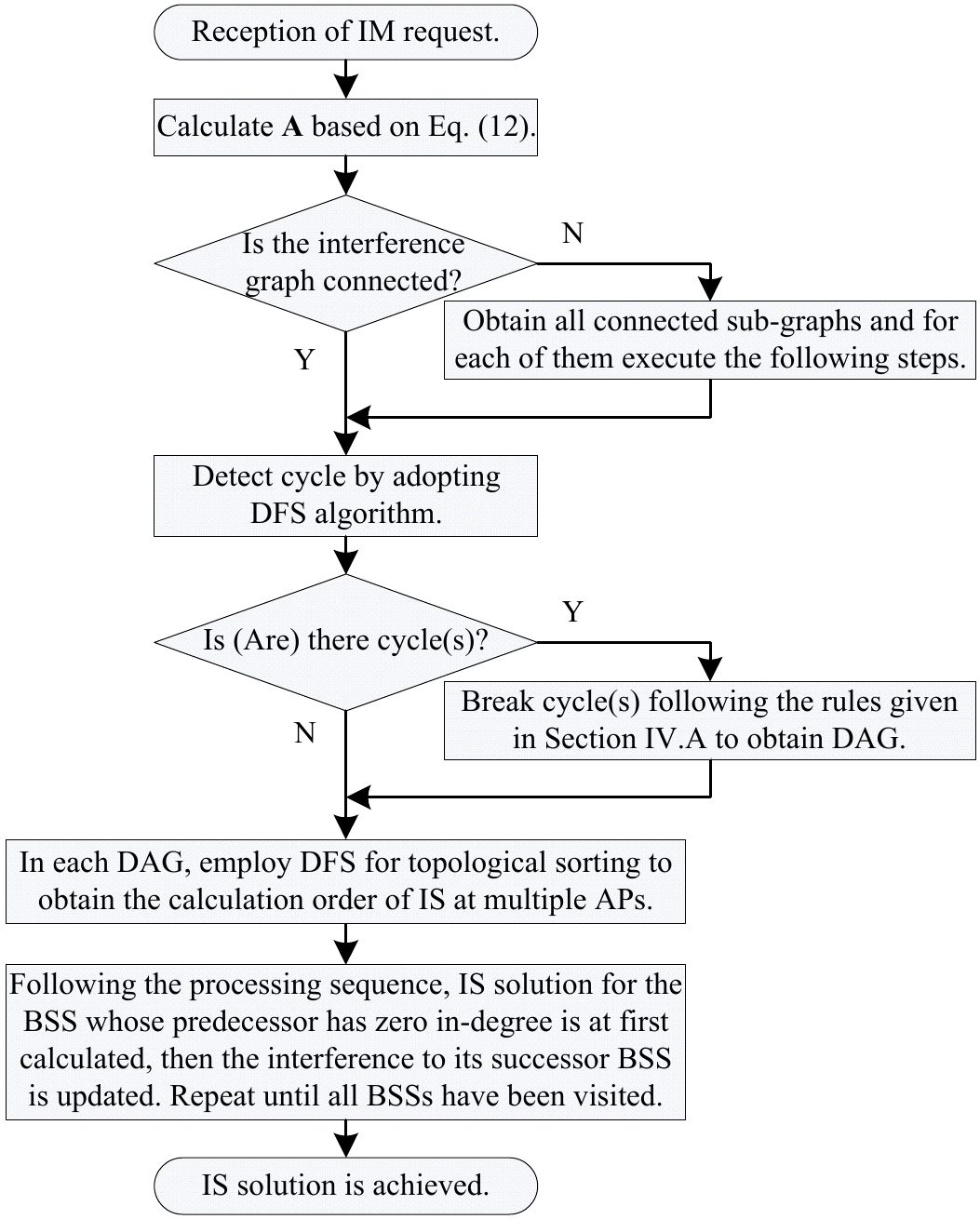}
%\vspace{-5pt}
\caption{Flowchart of IS implementation at the WLAN controller.}
\label{fig: figure eleven}
\vspace*{-0.05in}
\end{figure}

As for the WLAN controller, it maintains $\mathbf{C}$ and $\mathbf{T}$ based on
the information received from APs. Upon receiving an IM request from APs,
the controller calculates an IS solution following the procedure in Fig.~10.
Then, it sends instructions to each victim AP, based on which proper steering signals
are generated.
\setcounter{figure}{12}
\begin{figure*}[!htb]
%\vspace*{-0.05in}
\centering
\begin{minipage}{0.66\columnwidth}
\centering
\includegraphics[width=1.01\columnwidth, height = 1.72in]{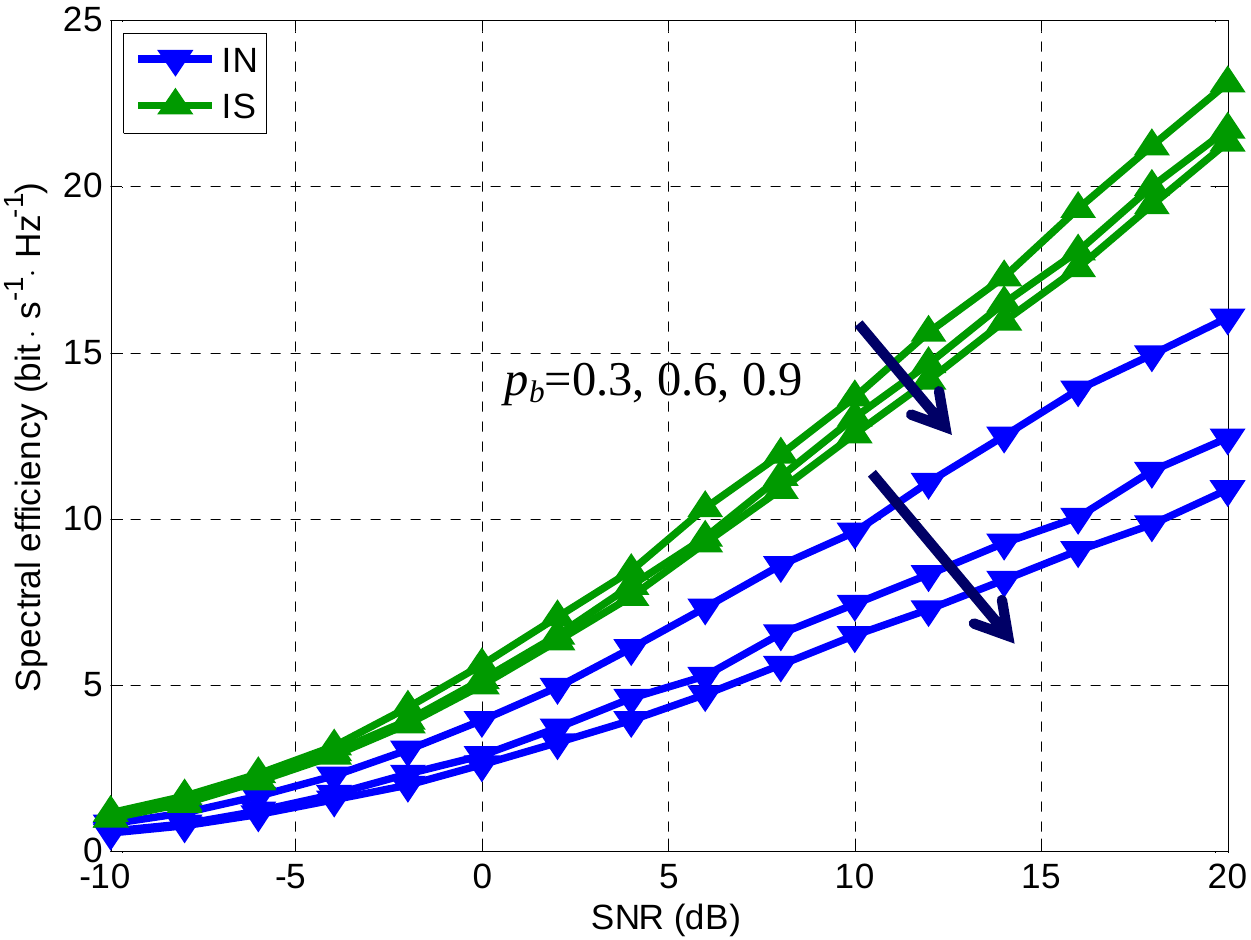}
\vspace{-14pt}
\caption{System SE under $K=3$, $p_b\in\{0.3, 0.6, 0.9\}$ and $\eta=2$.}
\label{fig: figure fourteen}
\end{minipage}
\hfill
\begin{minipage}{0.66\columnwidth}
\centering
\includegraphics[width=1.01\columnwidth, height = 1.72in]{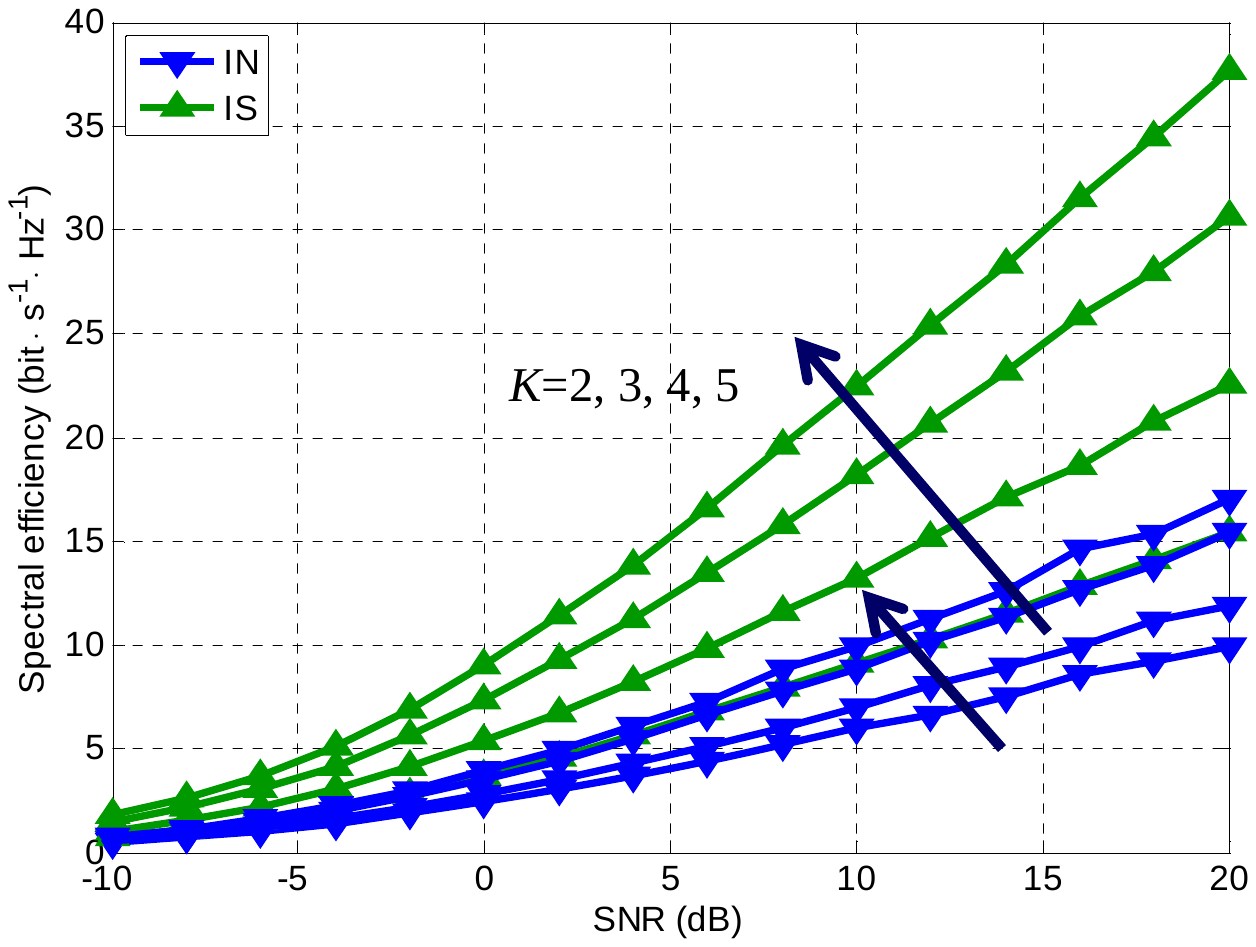}
\vspace{-14pt}
\caption{System SE under $p_b=0.9$, $K\in\{2, 3, 4, 5\}$ and $\eta=1$.}
\label{fig: figure fifteen}
\end{minipage}
\hfill
\begin{minipage}{0.66\columnwidth}
\centering
\includegraphics[width=1.01\columnwidth, height = 1.72in]{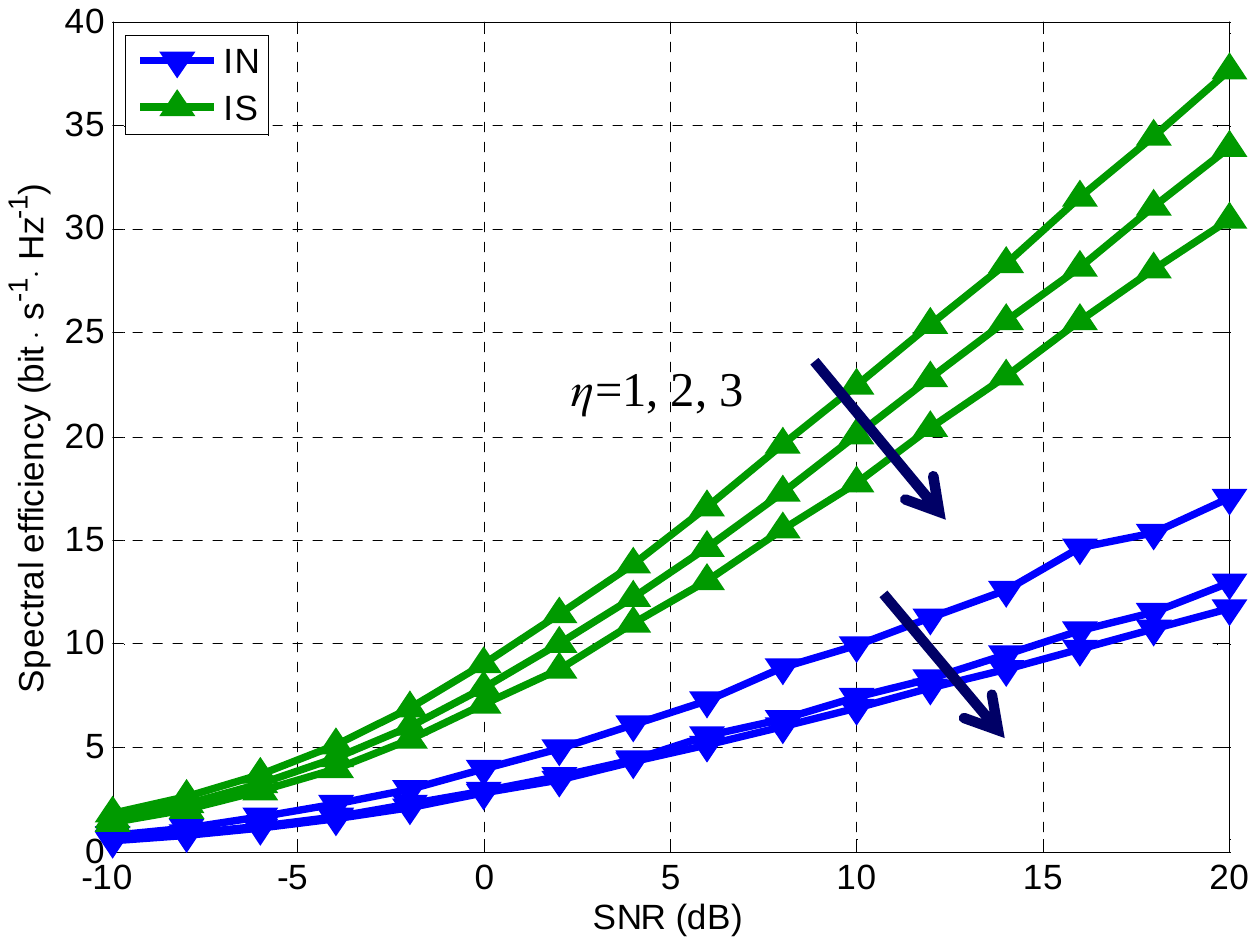}
\vspace{-14pt}
\caption{System SE under $K=5$, $p_b=0.9$ and $\eta\in\{1, 2, 3\}$.}
\label{fig: figure seventeen}
\end{minipage}
\vspace*{-0.2in}
\end{figure*}

According to the above discussion, IS can be implemented in WLANs with random network
topologies and arbitrary number of interferences.
Compared to existing WLAN protocols, the interferer is not required to do anything for IS
and only minor modifications are required at the victim AP and STA.
To reduce the scale of interference map and the corresponding processing complexity,
we can divide the entire WLAN into multiple sub-networks each of which consists of a limited
number of APs. IS is then realized in each part.
However, in such a case, interference between sub-networks cannot be managed.

\section{Evaluation}

We  evaluate the performance of IS using MATLAB simulation.
We set $N_{t}=N_{r}=2$. The system includes $K$ APs with random overlapping areas.
All APs have the same transmit power $P_{T}$.
We define the probability that the element in $\mathbf{A}$ is 1 as $p_b$.
By choosing different $p_b$s, an interference graph with various interference densities can be generated.
Since we focus on IS based on victim Tx and the advantages of IS over other non-victim-Tx based
implementations have been shown in Section III, here we only study the spectral efficiency of IS
and IN in a generalized WLAN with random number and distribution of interferences.
It should be noted that with IS or IN, a power overhead will be incurred at the executing AP;
when this power cost exceeds $P_{T}$, neither IS nor IN is applicable, thus yielding SE=0.
\setcounter{figure}{10}
\begin{figure}[!htb]
\vspace*{-0.05in}
%\graphicspath{/fig}
\centering
\includegraphics[width=0.34\textwidth,height=0.25\textwidth]{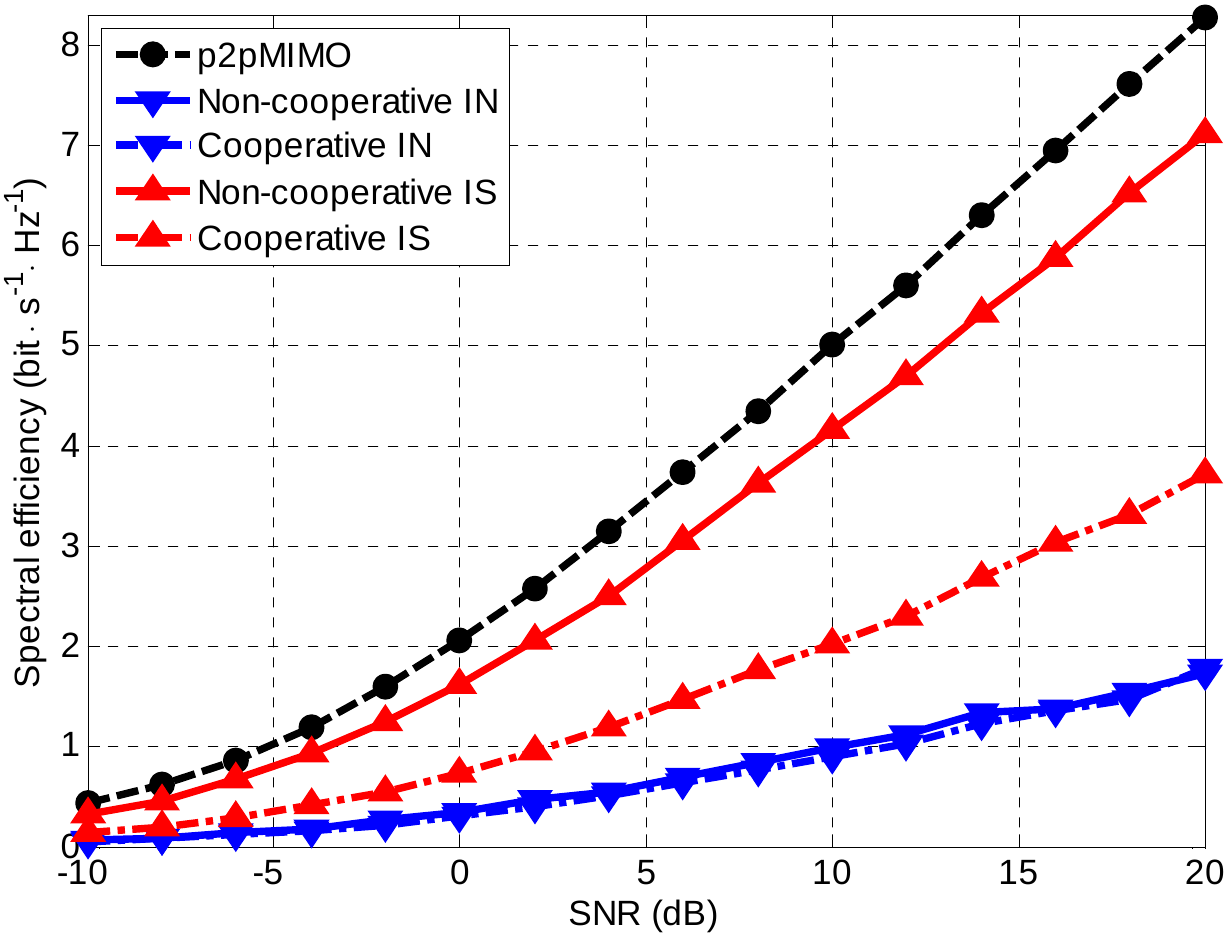}
\vspace*{-5pt}
\caption{Single transmission's SE with non-cooperative and cooperative IS/IN and single interference.}
\label{fig: figure twelve}
\vspace*{-0.05in}
\end{figure}

Fig.~11 plots the SE of a victim transmission in a BSS,
denoted by a vertex in an interference graph,
with non-cooperative and cooperative IS/IN and single interference.
IS is shown to outperform IN due to the reduced power overhead.
Cooperative IS is obviously inferior to non-cooperative IS,
whereas cooperative and non-cooperative IN yield statistically the same SE.
This is consistent with the results given in Fig.~8.
\begin{figure}[!htb]
\vspace*{-0.05in}
%\graphicspath{/fig}
\centering
\includegraphics[width=0.34\textwidth,height=0.25\textwidth]{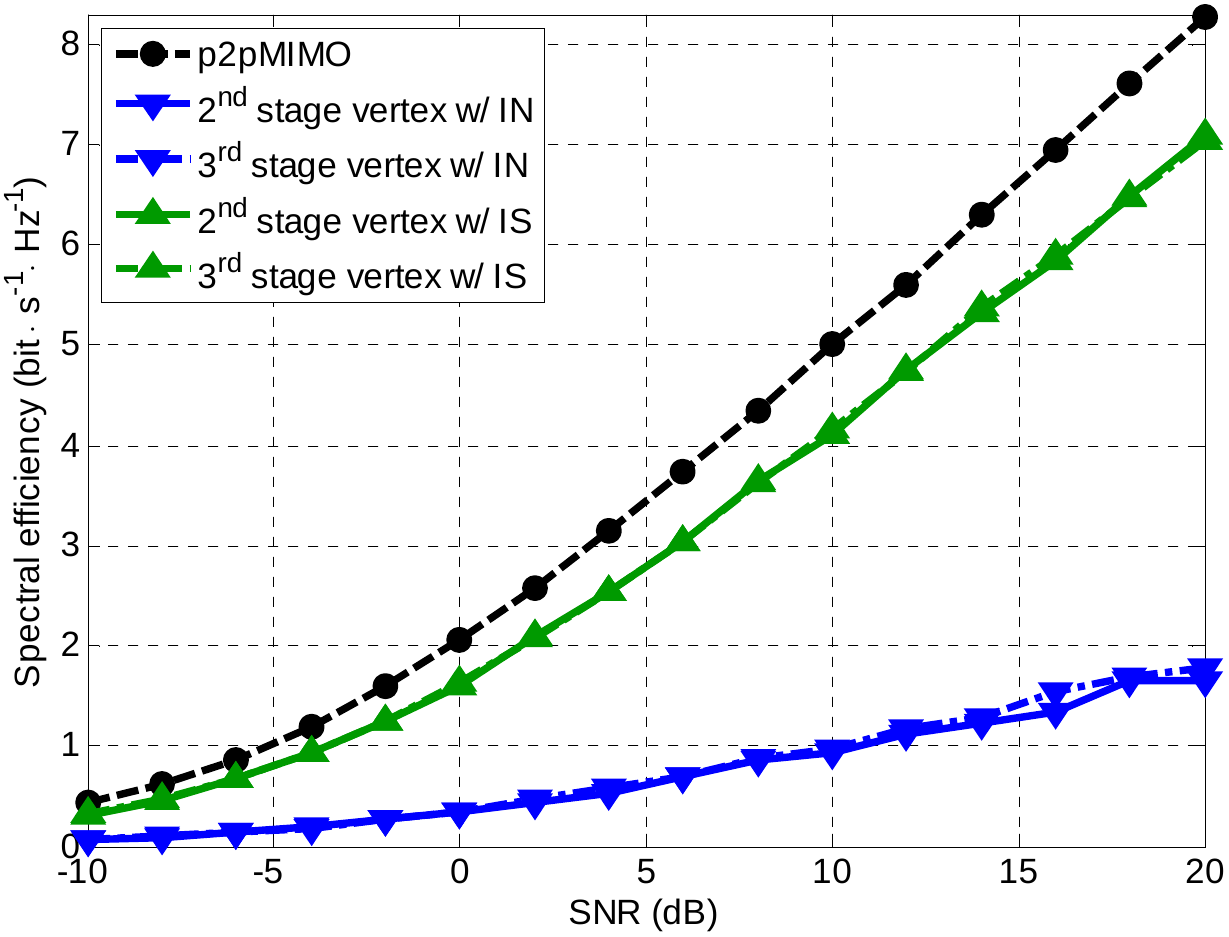}
\vspace*{-5pt}
\caption{Single transmission's SE at different processing stages.}
\label{fig: figure thirteen}
\vspace*{-0.05in}
\end{figure}

Fig.~12 shows the influence of processing stage on a victim user's average SE.
We adopt $K=3$ and the interference graph with a linear topology.
For example, AP$_1$ disturbs STA$_2$ associated with AP$_2$,
AP$_0$ interferes with STA$_1$ being served by AP$_1$,
and STA$_0$ associated with AP$_0$ is free of interference.
According to the IS implementation in Fig.~10,
since the indegree of $v_{0}$ is 0, i.e., interference-free,
the IS solution for its successor vertex $v_{1}$ should be calculated first,
and then $v_{2}$. We call $v_{1}$ and $v_{2}$ the $2^{nd}$ and $3^{rd}$ stage vertex, respectively.
It can be seen from Fig.~12 that the processing stage doesn't affect the victim's SE
under either IS or IN. This is because the victim's SE is dependent on the strength of interference
which can be referred to the result given in Fig.~7 where $MP_{T}$ is the total transmit power
of $M$ interferers, but in the given linear topology,
the strength of interference incurred to $v_1$ and $v_2$ is statically identical.
%\begin{figure}[!htb]
%\graphicspath{/fig}
%\centering
%\includegraphics[width=0.36\textwidth,height=0.27\textwidth]{fig/figure14.pdf}
%\vspace{-10pt}
%\caption{System SE under $K=3$, $p_b\in\{0.3, 0.6, 0.9\}$ and $\eta=2$.}
%\label{fig: figure fourteen}
%\vspace*{-0.1in}
%\end{figure}

Fig.~13 plots the average system SE with IS and IN under $K=3$ and various $p_b$s.
Each victim suffers from at most two interferences, i.e., $\eta=2$.
SE performance is shown to decrease as $p_b$ increases.
Given the same $p_b$, IS outperforms IN.
Moreover, since IS focuses on the mitigation of effective part of interference imposed on
the victim transmission, its SE is not as sensitive to $p_b$ as IN's.
%\begin{figure}[!htb]
%\graphicspath{/fig}
%\centering
%\includegraphics[width=0.36\textwidth,height=0.27\textwidth]{fig/figure15.pdf}
%\vspace{-10pt}
%\caption{System SE under $p_b=0.9$, $K\in\{2, 3, 4, 5\}$ and $\eta=1$.}
%\label{fig: figure fifteen}
%\vspace*{-0.1in}
%\end{figure}

Fig.~14 shows the average system SE with IS and IN
under fixed $p_b$, various $K$ and $\eta=1$.
Although given a fixed $p_b$, a larger $K$ yields higher CCI among BSSs,
since IS can effectively mitigate the influence of interference on each transmission,
system SE grows as $K$ increases.
Moreover, recall that IS requires much less power than IN,
more transmit power will be available for data transmission,
and hence IS outperforms IN in SE.
%\begin{figure}[!htb]
%\graphicspath{/fig}
%\centering
%\includegraphics[width=0.36\textwidth,height=0.27\textwidth]{fig/figure16.pdf}
%\vspace{-10pt}
%\caption{System SE under $K=5$, $p_b=0.9$ and $\eta\in\{1, 2, 3\}$.}
%\label{fig: figure sixteen}
%\vspace*{-0.1in}
%\end{figure}

Fig.~15 plots the average system SE with IS and IN under $K=5$, $p_b=0.9$,
and various $\eta$.
SE is shown to decrease with increasing $\eta$, because as $\eta$ grows, the aggregated
interference is strengthened and more power will be consumed for generating a steering signal.
Thus, each victim's SE decreases, degrading system SE.

\section{Conclusion}
In this paper, we have proposed and evaluated a novel interference
management  technique, called {\em{interference steering}} (IS).
By exploiting both CSI w.r.t.~and data carried in the interference(s),
a steering signal is generated so that the spatial feature of interference
is made orthogonal to the intended transmission at the victim receiver.
IS does not require any adjustment at the interferer and
only minor modifications are required at the victim transmission pair,
thus facilitating its practical implementation and deployment.
The proposed mechanism can be applied to general enterprise WLANs
with random network topologies and arbitrary number and distribution of interferences.
Our simulation results show that IS can significantly improve system SE over
the other existing IM methods.

%======================================================
%EE We delete this part on May 1st, 2016 %%START HERE%%
%======================================================
%Since IS consumes transmit power, how to balance this power cost with the IS executor's
%transmission performance so as to further improve network SE is an important issue
%that needs to be addressed.
%The RF signal used for IS may incur additional interference to other ongoing transmissions.
%Considering that the steering signal carries the same data as the interfering transmission in
%the single-interference case, harvesting this signal power may improve the interferer's transmission.
%These are matters of our future inquiry.
%======================================================
%EE We delete this part on May 1st, 2016   %%END HERE%%
%======================================================


\section*{Acknowledgments}
This work was supported in part by NSFC (61173135,~U14\\05255);
the 111 Project (B08038); the Fundamental Research Funds for the Central Universities (JB171503).
It was also supported in part by the US National Science Foundation under Grant 1317411.



%\bibliographystyle{IEEEtran}

\end{document}